\documentclass[a4paper,11pt]{article}
\pdfoutput=1

\usepackage{jheppub}

\usepackage[T1]{fontenc}

\usepackage{array}

\newcommand\myref{\refstepcounter{equation}\theequation}

\title{Wilson lines and gauge invariant off-shell amplitudes}

\author{Piotr Kotko}
\affiliation{The H. Niewodnicza\'nski Institute of Nuclear Physics,\\ Polish Academy of Sciences, \\ Radzikowskiego 152, 31-342 Krak\'ow, \\ Poland}
\emailAdd{piotr.kotko@ifj.edu.pl}

\arxivnumber{1403.4824}

\abstract{
We study matrix elements of Fourier-transformed straight infinite
Wilson lines as a way to calculate gauge invariant tree-level amplitudes
with off-shell gluons. 
The off-shell gluons  are assigned "polarization vectors"
which (in the Feynman gauge) are transverse to their off-shell momenta and define the direction of the corresponding Wilson line
operators.
The infinite Wilson lines are first regularized
to prove the correctness of the method. 
We have implemented the method in a computer FORM program that can
calculate gluonic matrix elements of Wilson line operators automatically.
In addition we formulate the Feynman rules that are convenient in
certain applications, e.g. proving the Ward identities. 
Using both the program and the Feynman rules
we calculate a few examples, in particular the matrix elements corresponding
to gauge invariant $g^{*}g^{*}g^{*}g$ and $g^{*}g^{*}g^{*}g^{*}g$ processes. An
immediate application of the approach is in the high energy scattering,
as in a special kinematic setup our results reduce to the form directly
related to Lipatov's vertices. Thus the results we present can be
directly transformed into Lipatov's vertices, in particular into $RRRP$
and $RRRRP$ vertices with arbitrary {}``orientation'' of reggeized
gluons. Since the formulation itself is not restricted to high-energy
scattering, we also apply the method to a decomposition of an ordinary
on-shell amplitude into a set of gauge invariant objects.
}

\begin{document} 
\maketitle
\flushbottom

\section{Introduction}

A basic off-shell object in QCD is a momentum space Green's function,
i.e. the Fourier transform of a matrix element of time ordered field
operators (in what follows we will consider only gluon operators for
definiteness). In case of the most standard collinear factorization
(see e.g.\ \cite{Collins:2011zzd} for a review) the hard amplitude
defining the perturbative core of the process is defined by the
reduced connected Green's function, i.e. the external propagators
are amputated, taken on-shell and contracted with polarization vectors.
Thanks to on-shellness and transversality of the polarization vectors
to corresponding momenta such amplitudes are gauge invariant. However,
for processes occurring at high energies one often has to deal with
so called high-energy (or $k_{T}$) factorization \cite{Gribov:1984tu,Catani:1990eg,Collins:1991ty,Catani:1994sq}
(we put aside here the issues concerning the factorization breaking,
see e.g.\ \cite{Xiao:2010sp,Dominguez:2011wm,Mulders:2011zt,Avsar:2011tz,Avsar:2012hj}
for more details). In that case the reduction of the Green's function
does not put all of the external legs on-shell; the remaining one
or two off-shell legs are contracted with eikonal vectors corresponding
to fast moving hadrons and off-shell momenta are restricted to be
transverse to the pertinent eikonal vectors. There is however an issue
related to gauge invariance of such objects. In general, in order
to maintain the gauge invariance additional non-standard (i.e. not
calculable from standard QCD Feynman rules) contributions are needed.
One of the approaches is to use the Lipatov's effective action \cite{Lipatov:1995pn,Antonov:2004hh}
and interpret an off-shell gluon with additional contributions as
an effective reggeized gluon $R$. Even at tree-level this approach
is rather complicated for multiple final states. Therefore only recently
some automatic methods to calculate such amplitudes for larger multiplicities
have been developed \cite{vanHameren:2012if,vanHameren:2012uj,vanHameren:2013csa}
(examples of practical applications were presented in \cite{vanHameren:2013fla,vanHameren:2014lna}).
They use different methods than the Lipatov's effective action (see also e.g. \cite{Leonidov:1999nc} for yet another approach).
In particular, the  Lipatov's effective action uses Wilson lines, i.e.
path ordered exponentials of color gauge fields, while the other methods do not refer
to Wilson lines directly. However, the Wilson lines are viewed as the
basic objects at very high energies (see e.g.\ \cite{Balitsky:2001gj} or the Color Glass Condensate formulation
of QCD \cite{Gelis:2010nm}), therefore they are always present
in one or the other form. For instance in \cite{vanHameren:2012if} they show up as eikonalized quarks.
 Actually, as we will see in the present
paper also the additional contributions recovering the gauge invariance constructed in ref.~\cite{vanHameren:2012uj} from the Slavnov-Taylor identities
 do correspond to a bremsstrahlung
from a straight infinite Wilson line. Basing on this observation we will
formulate a prescription to calculate off-shell gauge-invariant
``amplitudes'' by considering matrix elements of Fourier transforms
of straight infinite Wilson line operators. The ``momentum'' of
such an operator corresponds to an off-shell gluon (and additional
contributions needed by the gauge invariance), while the direction
of the Wilson line corresponds to its {}``polarization'' vector.
In our prescription the momenta and direction of a Wilson line are
arbitrary, except that they have to be mutually transverse. This allows
to apply the method also outside the high energy factorization approach
as we will see.

In order to test the method we have implemented it in a computer program
written in FORM \cite{Kuipers:2012rf}. It allows to calculate matrix
elements of Wilson line operators for several external legs analytically.
For instance, using the program a calculation of a process with four
reggeized gluons with arbitrary {}``orientation'' and an additional
gluon emission, $RRRRg$, can be done automatically.

Let us collect at this point the main elements of the paper. i) Any
tree-level amplitude with arbitrary number of gluonic off-shell legs
and any number of on-shell legs, where the off-shell gluons have polarization
vectors transverse to their off-shell momenta, can be made gauge invariant
by assigning a proper infinite Wilson line operators to off-shell
gluons. Those operators are at first sight ill-defined and we develop
their regularized version to prove the correctness of the approach.
To this end we also prove the Ward identities. ii) The off-shell amplitudes
we consider here are more general than the ones appearing in the high-energy
literature, but they reduce to the Lipatov's vertices with certain
choice of the off-shell momenta and Wilson line directions (some of
the contributions vanish with that choice). We check some explicit
examples using the Feynman diagrams, in particular we give an example
for the gauge invariant $g^{*}g^{*}g^{*}g$ process. iii) We construct
a computer program that can calculate off-shell gluonic amplitudes
automatically and analytically, using the presented method. Using
the program we cross-check the result for $g^{*}g^{*}g^{*}g$ and
calculate the gauge invariant $g^{*}g^{*}g^{*}g^{*}g$ matrix element.

The work is organized as follows. The first two sections are in a
sense introductory. In section \ref{sec:gauge_restoring_amp} we will
introduce Wilson lines in the context of the off-shell amplitudes.
We choose to do this by taking as an example the result of ref.~\cite{vanHameren:2012uj}
for off-shell high energy amplitudes. In section \ref{sec:basics_of_gaugelink}
we recall some basics concerning Wilson lines. Next, in section \ref{sec:formal_develop}
we make some more formal definitions of the off-shell amplitudes using
the Wilson lines. In section \ref{sec:OGIME} we shortly present the
computer program based on the method. We introduce the Feynman rules
in section \ref{sec:Feynman_rules} and prove the Ward identities
in section \ref{sec:Word_ident}. In section \ref{sec:Examples} we
give some examples of explicit calculations. Next, in section \ref{sec:GIdecomp},
we present a potentially interesting application of the present approach
in decomposing ordinary amplitudes into gauge invariant pieces. Finally,
we make some summarizing remarks in section \ref{sec:summary}.

\section{High energy amplitudes and gauge invariance}

\label{sec:gauge_restoring_amp}

In order to introduce the Wilson lines in the context of off-shell
amplitudes, let us start with a short recollection of the high-energy
factorization of Catani, Ciafaloni and Hautmann (CCH) \cite{Catani:1990eg,Catani:1994sq}.
For more detailed albeit compact review we refer e.g. to \cite{vanHameren:2012uj,vanHameren:2013fla}.
In the original CCH approach a hadro- and lepto-production of heavy
quarks was considered. At high energies, the relevant hard partonic
sub-amplitudes turn out to be off-shell, i.e. we have to consider
amplitudes $g^{*}g^{*}\rightarrow Q\overline{Q}$ or $\gamma g^{*}\rightarrow Q\overline{Q}$.
They are defined by the Green's function with the on-shell legs amputated,
while the off-shell gluon legs (including propagators) are contracted
with so called eikonal vertices. To be more precise, if the momentum
of the hadron $A$ is $p_{A}$ the corresponding eikonal vertex is
just $p_{A}^{\mu}$ (modulo a prefactor). Moreover, the momentum $k_{A}$
of the corresponding off-shell leg has the form $x_{A}p_{A}+k_{TA}$,
i.e. it is transverse to $p_{A}$ ($k_{T}\cdot p_{A}=0$, $p_{A}^{2}=0$).
Since the CCH factorization is stated in the axial gauge, it turns
out that the standard Feynman diagrams are enough to obtain the gauge
invariant set of diagrams (recall we consider the heavy quark production
case here). Moreover, for heavy quark lepto-production it is even true
for any additional radiation of gluons. Therefore in \cite{Catani:1994sq}
the CCH factorization was demonstrated to hold up to several loops
for DIS heavy quark structure function. The last statement is also
true for the so-called hybrid version of CCH factorization in hadron-hadron
collision, i.e. where only one gluon is off-shell \cite{Deak:2009xt,vanHameren:2012uj}.
This approach is thought to be a good approximation to the full high-energy
factorization in case of forward processes, e.g.\ forward jet production.
However in the case of jets one usually needs also purely gluonic
sub-processes. In that case the off-shell sub-process $gg^{*}\rightarrow g\ldots g$
is not gauge invariant. One can however still get the correct result
with a particular choice of polarization vectors, but some modern
methods relying on gauge invariance (e.g.\  the helicity method)
cannot be used in that case. The most natural approach to the gauge
invariance problem in the above context is just to see what kind of
terms violate the gauge invariance and try to make use of that knowledge.
This path was taken in ref.~\cite{vanHameren:2012uj} and now we
shall briefly recall this method and point out the connection to the
Wilson line.

The high-energy gluonic amplitude $\mathcal{M}_{p_{A}}\left(\varepsilon_{1},\ldots,\varepsilon_{N}\right)$
with a single off-shell leg with incoming momentum $k_{A}$ is defined
by the following reduction formula \begin{multline}
\mathcal{M}_{p_{A}}\left(\varepsilon_{1},\ldots,\varepsilon_{N}\right)=\lim_{k_{A}\cdot p_{A}\rightarrow0}\,\lim_{k_{1}^{2}\rightarrow0}\ldots\lim_{k_{N}^{2}\rightarrow0}\,\, p_{A}^{\mu_{A}}\, k_{1}^{2}\varepsilon_{1}^{\mu_{1}}\ldots k_{N}^{2}\varepsilon_{N}^{\mu_{N}}\\
\,\tilde{G}_{\mu_{A}\mu_{1}\ldots\mu_{N}}\left(k_{A},k_{1},\ldots,k_{N}\right),\label{eq:Reduction}\end{multline}
where $\varepsilon_{1},\ldots,\varepsilon_{N}$ are polarization vectors
of on-shell gluons with momenta $k_{1},\ldots,k_{N}$ and $\tilde{G}$
is the momentum space Green's function. The internal (off-shell) propagators
of $\tilde{G}$, including the leg with off-shell momentum $k_{A}$
are taken to be in the axial gauge with the gauge vector $p_{A}$,
whereas the legs with momenta $k_{1},\ldots,k_{N}$ are in the Feynman
gauge. This is allowed as it is known that for the legs that are eventually
on-shell one can choose a different gauge than for internal lines.
Thanks to the first limit in (\ref{eq:Reduction}) the momentum $k_{A}$
has the structure\begin{equation}
k_{A}^{\mu}=z_{A}\, p_{A}^{\mu}+k_{TA}^{\mu}\label{eq:kA}\end{equation}
with $k_{A}^{2}=-\left|\vec{k}_{T\, A}\right|^{2}$. The off-shell
momentum $k_{A}$ has thus the structure complementary to the high-energy
factorization described above%
\footnote{There is a difference with respect to a definition of the corresponding
amplitude given in \cite{vanHameren:2012uj}, where additional factors
were inserted to maintain collinear limit already at this stage. The
relation is $\mathcal{M}=\left|\vec{k}_{T\, A}\right|z_{A}\,\mathcal{M}_{p_{A}}$,
where l.h.s is the amplitude of \cite{vanHameren:2012uj}.%
}. As already mentioned the amplitude $\mathcal{M}_{p_{A}}$ is in
general not gauge invariant in the sense of the following Ward identity
(unless we choose the polarization vectors in a special way)\begin{equation}
\mathcal{M}_{p_{A}}\left(\ldots,k_{i},\ldots\right)\neq0.\label{eq:Ward_viol}\end{equation}
However, since we have the relation between the Green's function and
$\mathcal{M}_{p_{A}}$ we can actually calculate the r.h.s. of (\ref{eq:Ward_viol})
using the Slavnov-Taylor identities (for an elementary review of the
Slavnov-Taylor identities see e.g.\ \cite{Arodz:2010}). Furthermore,
it turns out that within the gauge we are using the sum of all the
gauge contributions with a proper treatment of external ghosts gives
a {}``gauge-restoring amplitude'' $\mathcal{W}$, such that $\tilde{\mathcal{M}}_{p_{A}}=\mathcal{M}_{p_{A}}+\mathcal{W}$
satisfies the Ward identity \begin{equation}
\tilde{\mathcal{M}}_{p_{A}}\left(\ldots,k_{i},\ldots\right)=0.\end{equation}
In order to write the amplitude $\mathcal{W}$ in a compact manner,
let us recall that any gluonic amplitude $\mathcal{M}$ may be decomposed
into so called color-ordered amplitudes \cite{Mangano:1987xk} as
follows (we omit the polarization vectors here)

\begin{equation}
\mathcal{M}=\sum_{\Pi'\left(a_{1},\ldots,a_{M}\right)}\mathrm{Tr}\left(t^{a_{1}}\ldots t^{a_{M}}\right)\,\mathcal{M}^{\left(a_{1}\ldots a_{M}\right)},\label{eq:Color_ord_def}\end{equation}
where $a_{1},\ldots,a_{M}$ are color indices of the gluons with momenta
$k_{1},\ldots,k_{M}$ respectively, $\Pi'$ is the set of all non-cyclic
permutations of the color indices, the matrices $t^{a}$ are generators
of the $\mathrm{SU}\left(3\right)$ color group normalized as $\mathrm{Tr}\left(t^{a}t^{b}\right)=\delta^{ab}/2$.
The amplitude $\mathcal{M}^{\left(a_{1}\ldots a_{N}\right)}$ corresponds
to a particular color ordering given in the superscript (see \cite{Mangano:1990by}
for a review of color-ordering techniques). Specifically, in the present
case the color decomposition for the off-shell amplitude reads

\begin{equation}
\tilde{\mathcal{M}}_{p_{A}}=\sum_{\Pi'\left(a_{A},a_{1},\ldots,a_{N}\right)}\mathrm{Tr}\left(t^{a_{A}}t^{a_{1}}\ldots t^{a_{N}}\right)\,\tilde{\mathcal{M}}_{p_{A}}^{\left(a_{A}a_{1}\ldots a_{N}\right)}.\label{eq:Color_ord}\end{equation}
The result for the color-ordered version of the {}``gauge-restoring
amplitude'' $\mathcal{W}$ turns out to be very simple\begin{multline}
\mathcal{W}^{\left(a_{A}a_{1}\ldots a_{N}\right)}\left(\varepsilon_{1},\ldots,\varepsilon_{N}\right)=-\left|\vec{k}_{T\, A}\right|\left(\frac{-g}{\sqrt{2}}\right)^{N-1}\\
\frac{\,\varepsilon_{1}\cdot p_{A}\ldots\varepsilon_{N}\cdot p_{A}}{k_{1}\cdot p_{A}\left(k_{1}-k_{2}\right)\cdot p_{A}\ldots\left(k_{1}-k_{2}-\ldots-k_{N-1}\right)\cdot p_{A}}.\label{eq:W}\end{multline}

\begin{figure}
\begin{centering}
\includegraphics[width=0.4\textwidth]{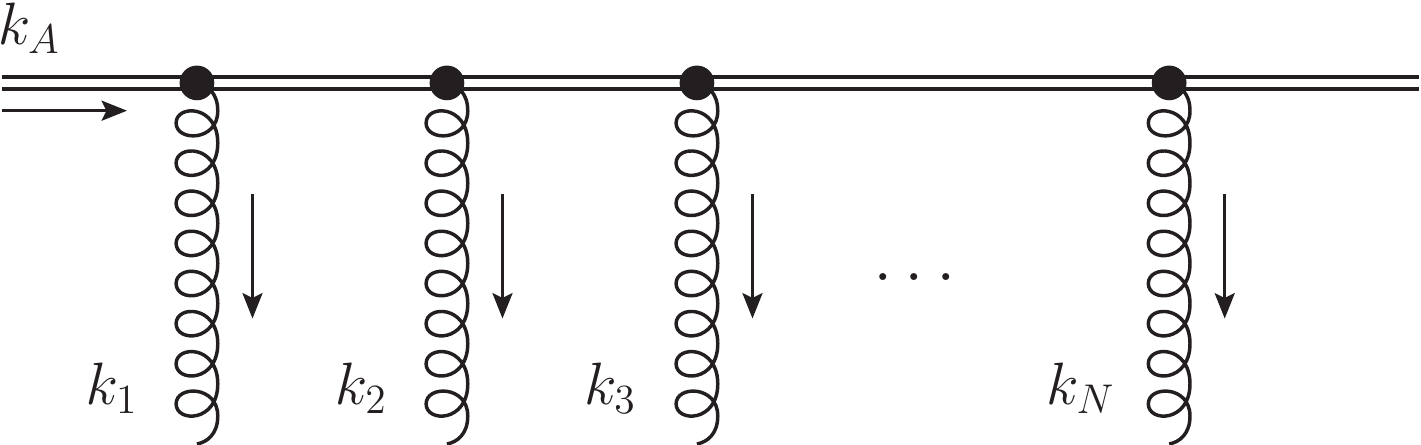}
\par\end{centering}

\caption{\label{fig:W_as_Wilsonline}\small Graphical representation of the
{}``gauge-restoring amplitude'' $\mathcal{W}$. Each gluon is coupled
via $p_{A}^{\mu}$ thus giving a factor $p_{A}\cdot\varepsilon_{i}$
for a gluon with momentum $k_{i}$ and polarization vector $\varepsilon_{i}$.
The double-line propagator between subsequent emissions has the form
$i/k\cdot p_{A}$ where $k$ is a momentum of the line. This picture
will be further interpreted as a gauge link.}
\end{figure}

The above result (\ref{eq:W}) has a very transparent structure and
expresses certain bremsstrahlung contributions. Looking at the numerator,
we see that each of the external on-shell gluons is coupled directly
to the eikonal vector $p_{A}$. There are no triple or quartic gluon
vertices as we have chosen the axial gauge with $p_{A}$ as the gauge
vector; the propagator is thus always perpendicular to $p_{A}$ and
all such couplings are eliminated. Next, looking at the denominator,
we see that there are certain scalar propagators between each emission,
i.e. we have $i/k\cdot p_{A}$ between each emission with $k$ being
the momentum remaining after the last emission. This is illustrated
in figure~\ref{fig:W_as_Wilsonline}. 

For people working with the collinear factorization the eikonal couplings
and eikonal propagators are mainly familiar from the Feynman rules
for PDFs \cite{Collins:1981uw,Collins:2011zzd}. They originate in
a straight Wilson line connecting two fields separated on the light
cone (and making the whole object gauge invariant). This may suggest,
that the high-energy amplitude $\tilde{\mathcal{M}}_{p_{A}}$ is also
related to a straight Wilson line. Indeed, as we shall see below the
straight infinite Wilson line has a structure complementary with the
structure of $\mathcal{W}$.

\section{Basics of Wilson lines}

\label{sec:basics_of_gaugelink}

Let us now recall some basic facts regarding gauge links (or Wilson
lines; we shall use both terms interchangeably) and define our notation.
For a more comprehensive review of Wilson lines in the context of
Quantum Field Theory we refer e.g.\ to \cite{Arodz:2010} (in that
reference a gauge link is referred to as the {}``parallel-transporter''
due to its geometrical interpretation).

The Wilson line along the path $\mathcal{C}$ joining two space-time
points $x,y$ is defined as\begin{equation}
\left[x,y\right]_{\mathcal{C}}=\mathcal{P}\exp\left\{ ig\,\int_{C}dz_{\mu}A_{b}^{\mu}\left(z\right)t^{b}\right\} ,\label{eq:gaugelink_def}\end{equation}
where $\mathcal{P}$ is the operation of ordering color matrices $t^{b}$
along the path. Its crucial property is the transformation law under
the local gauge transformations $U\left(x\right)$\begin{equation}
\left[x,y\right]_{\mathcal{C}}\rightarrow U\left(x\right)\left[x,y\right]_{\mathcal{C}}U^{\dagger}\left(y\right).\label{eq:gaugelink_transf}\end{equation}

In many applications it is convenient to choose a straight line as
the path $\mathcal{C}$. Defining the straight line to lie along the
direction specified by the four vector $n$, it can be parametrized
as\begin{equation}
z^{\mu}\left(s\right)=x^{\mu}+sn^{\mu},\quad z^{\mu}\left(1\right)=y^{\mu}.\end{equation}
Then for a gauge link from $x$ to $y$ we have simply\begin{equation}
\left[x,y\right]_{n}=\mathcal{P}\exp\left\{ ig\,\int_{0}^{1}ds\, n_{\mu}A_{b}^{\mu}\left(x+sn\right)t^{b}\right\} .\label{eq:gaugelink_def-1}\end{equation}
Note, that we have used a subscript $n$ to denote the path direction.

In the present paper we shall use the paths extending from minus to
plus infinity. Let us introduce the following formal definition\begin{equation}
\left[x\right]_{n}\equiv\mathcal{P}\exp\left\{ ig\,\int_{-\infty}^{\infty}ds\, n\cdot A_{b}\left(x+sn\right)t^{b}\right\} .\label{eq:gaugelink_def-1-1}\end{equation}
The subtleties concerning the (divergent) integration over $ds$ will
be discussed below in section~\ref{sec:formal_develop}. Consider
now the expansion of the gauge link defined above:\begin{multline}
\left[x\right]_{n}=\mathcal{P}\Big\{1+ig\,\int_{-\infty}^{\infty}ds\, n\cdot A_{b}\left(x+sn\right)t^{b}\\
+\left(ig\right)^{2}\frac{1}{2!}\,\int_{-\infty}^{\infty}ds\,\int_{-\infty}^{\infty}ds'\, n\cdot A_{b}\left(x+sn\right)\, n\cdot A_{b'}\left(x+s'n\right)t^{b}t^{b'}+\ldots\Big\}.\end{multline}
In order to utilize the path ordering we use the symmetry of the integrands
with respect to $s,s',\ldots$ and obtain \begin{multline}
\left[x\right]_{n}=1+ig\,\int_{-\infty}^{\infty}ds\, n\cdot A_{b}\left(x+sn\right)t^{b}\\
+\left(ig\right)^{2}\int_{-\infty}^{\infty}ds\int_{-\infty}^{s}ds'\, n\cdot A_{b}\left(x+sn\right)\, n\cdot A_{b'}\left(x+s'n\right)t^{b}t^{b'}+\ldots.\label{eq:gaugelink_[x]n}\end{multline}
Note, that the factorials originating from the expansion of the exponential
cancel in the formula above.

\section{Formal developments}

\label{sec:formal_develop}

Let us now make a more formal connection of (\ref{eq:W}) to a gauge
link. At this point we see from (\ref{eq:gaugelink_[x]n}), that if
the Wilson line direction is defined via the four vector $p_{A}$
appearing in (\ref{eq:kA}), the external gluons will be contracted
with $p_{A}$, as desired in view of eq.\ (\ref{eq:W}). The eikonal
propagators in (\ref{eq:W}) will come out from the path ordered integrals.
In the following we make those statements more formal.

First, note that the gauge link $\left[x\right]_{n}$ is gauge invariant
on its own. This is due to eq.~(\ref{eq:gaugelink_transf}) and the
fact that for a local gauge transformation $U\left(x\right)$ (more
precisely for a small local gauge transformation) we have $U\left(x\right)\rightarrow1$
for $\left|\vec{x}\right|\rightarrow\infty$. So $\left[x\right]_{n}$
is gauge invariant for any $x$ as long as $n$ has a nonzero spatial
component. For definiteness we may restrict $n$ to be non-light-like
at this point, but the gauge invariance of the matrix elements defined
below will be maintained for any $n$.

Consider next a matrix element of an operator defined as follows\begin{equation}
\mathfrak{R}_{n}^{c}\left(k\right)=\int d^{4}x\, e^{ix\cdot k}\mathrm{Tr}\left\{ \frac{1}{\pi g}\, t^{c}\left[x\right]_{n}\right\} .\label{eq:R_def}\end{equation}
More precisely, we consider \begin{equation}
\mathfrak{M}\left(n,\varepsilon_{1},\ldots,\varepsilon_{N}\right)\overset{*}{=}\left\langle k_{1},\varepsilon_{1},c_{1};\ldots;k_{N},\varepsilon_{N},c_{N}\right|\mathfrak{R}_{n}^{c}\left(k\right)\left|0\right\rangle ,\label{eq:ME_def}\end{equation}
where $\left|k_{i},\varepsilon_{i},c_{i}\right\rangle $ is an external
on-shell gluon state in the Heisenberg picture with momentum $k_{i}$,
polarization vector $\varepsilon_{i}$ and color $c_{i}$. The star
adorning the equality sign means that only connected contributions
(i.e. proportional to $\delta^{4}\left(k_{A}+k_{1}+\ldots+k_{N}\right)$)
are to be taken into account. At this stage, the momentum $k$ is
arbitrary, i.e. it is not restricted to the form similar to (\ref{eq:kA}).
However, we shall see below in section \ref{sec:Feynman_rules} that
the form of the operator (\ref{eq:R_def}) assures that its matrix
element is proportional to $\delta\left(k\cdot n\right)$ implying
thus the high-energy kinematics (\ref{eq:kA}) if $n=p_{A}$. 

The matrix element (\ref{eq:ME_def}) can be calculated explicitly
in perturbation theory using field operators and the Wick theorem.
The only complication is due to the integrals over the path parameters
in (\ref{eq:gaugelink_[x]n}) which are formally divergent. In order
to define a suitable prescription, we propose the following finite-length
(regularized) version of $\left[x\right]_{n}$ \begin{equation}
\left[x\right]_{n}^{\left(\epsilon\right)}\equiv\left[x-\frac{2}{\epsilon}n,x+\frac{2}{\epsilon}n\right],\label{eq:[x]reg1}\end{equation}
with the path defined as\begin{equation}
z_{\epsilon}^{\mu}\left(s\right)=x^{\mu}+\frac{2}{\epsilon}\tanh\left(\frac{\epsilon s}{2}\right)\, n^{\mu},\,\,\,\,\, s\in\left(-\infty,\infty\right).\label{eq:[x]reg2}\end{equation}
Note that\begin{equation}
z_{\epsilon}^{\mu}\left(s\right)=x^{\mu}+sn^{\mu}+\mathcal{O}\left(\epsilon^{2}\right)\label{eq:[x]reg2a}\end{equation}
thus in the $\epsilon\rightarrow0$ limit we recover $\left[x\right]_{n}$.
Now, the regularized version of the expansion (\ref{eq:gaugelink_[x]n})
reads\begin{multline}
\left[x\right]_{n}^{\left(\epsilon\right)}=1+ig\,\int_{-\infty}^{\infty}ds\,\mathrm{sech}^{2}\lambda_{s}\, n\cdot A_{b}\left(z_{\epsilon}\left(s\right)\right)t^{b}\\
+\left(ig\right)^{2}\int_{-\infty}^{\infty}ds\int_{-\infty}^{s}ds'\,\mathrm{sech}^{2}\lambda_{s}\,\mathrm{sech}^{2}\lambda_{s'}\, n\cdot A_{b}\left(z_{\epsilon}\left(s\right)\right)\, n\cdot A_{b'}\left(z_{\epsilon}\left(s'\right)\right)t^{b}t^{b'}+\ldots,\label{eq:[x]reg3}\end{multline}
where\begin{equation}
\lambda_{s}=\frac{\epsilon s}{2}.\end{equation}
Let us pass to the momentum space\begin{multline}
\left[x\right]_{n}^{\left(\epsilon\right)}=1+ig\,\int\frac{d^{4}p}{\left(2\pi\right)^{4}}\, e^{-ip\cdot x}n\cdot\tilde{A}_{b}\left(p\right)\int_{-\infty}^{\infty}ds\,\mathrm{sech}^{2}\lambda_{s}\, e^{-i\frac{2}{\epsilon}\tanh\lambda_{s}\, p\cdot n}\, t^{b}\\
+\left(ig\right)^{2}\int\frac{d^{4}p}{\left(2\pi\right)^{4}}\frac{d^{4}p'}{\left(2\pi\right)^{4}}\, e^{-i\left(p+p'\right)\cdot x}n\cdot\tilde{A}_{b}\left(p\right)\, n\cdot\tilde{A}_{b'}\left(p'\right)\\
\times\int_{-\infty}^{\infty}ds\int_{-\infty}^{s}ds'\,\mathrm{sech}^{2}\lambda_{s}\,\mathrm{sech}^{2}\lambda_{s'}e^{-i\frac{2}{\epsilon}\tanh\lambda_{s}\, p\cdot n}e^{-i\frac{2}{\epsilon}\tanh\lambda_{s'}\, p'\cdot n}\, t^{b}t^{b'}+\ldots,\label{eq:[x]reg4-1}\end{multline}
where $\tilde{A}$ is the Fourier-transformed gauge field. There are
two types of integrals. The first one,\begin{equation}
I_{\epsilon}\left(p\cdot n\right)=\int_{-\infty}^{\infty}ds\,\mathrm{sech}^{2}\lambda_{s}\, e^{-i\frac{2}{\epsilon}\tanh\lambda_{s}\, p\cdot n}=\frac{2}{p\cdot n}\sin\left(\frac{2p\cdot n}{\epsilon}\right)\equiv\delta_{\epsilon}\left(p\cdot n\right),\label{eq:delta_repr}\end{equation}
is an approximation to the Dirac delta function,\begin{equation}
\lim_{\epsilon\rightarrow0}\delta_{\epsilon}\left(x\right)=2\pi\delta\left(x\right),\end{equation}
since \begin{equation}
\int dx\,\delta_{\epsilon}\left(x\right)=2\pi.\end{equation}
The second integral is\begin{equation}
J_{\epsilon}\left(s,p\cdot n\right)=\int_{-\infty}^{s}ds'\,\mathrm{sech}^{2}\lambda_{s'}\, e^{-i\frac{2}{\epsilon}\tanh\lambda_{s'}\, p\cdot n}=\frac{i}{p\cdot n}\,\left(e^{-i\frac{2}{\epsilon}\tanh\lambda_{s}\, p\cdot n}-e^{i\frac{2}{\epsilon}\, p\cdot n}\right).\label{eq:Ie2}\end{equation}
Those two types of integrals propagate for any of the terms in the
expansion (\ref{eq:[x]reg4-1}). The second exponent in the r.h.s.
of (\ref{eq:Ie2}) will not contribute to the result in the $\epsilon\rightarrow0$
limit. This is because it oscillates rapidly and its contribution
is zero due to the Riemann\textendash{}Lebesgue lemma. Therefore for
the regularized version of the operator (\ref{eq:R_def}) we obtain\begin{multline}
\mathfrak{R}_{n}^{c\,\left(\epsilon\right)}\left(k\right)=i\, n\cdot\tilde{A}_{b}\left(k\right)\frac{\delta_{\epsilon}\left(k\cdot n\right)}{2\pi}\,2\mathrm{Tr}\left(t^{c}t^{b}\right)\\
+i^{2}g\int\frac{d^{4}p}{\left(2\pi\right)^{4}}\, n\cdot\tilde{A}_{b}\left(p\right)\, n\cdot\tilde{A}_{b'}\left(k-p\right)\\
\times\frac{i}{p\cdot n}\,\left[\frac{\delta_{\epsilon}\left(k\cdot n\right)}{2\pi}+\mathcal{O}\left(\epsilon\right)\right]\,2\mathrm{Tr}\left(t^{c}t^{b}t^{b'}\right)+\ldots,\label{eq:[x]reg4-1-1}\end{multline}
where the $\mathcal{O}\left(\epsilon\right)$ terms come from the
second exponent in (\ref{eq:Ie2}). In the limit $\epsilon\rightarrow0$
we get\begin{multline}
\mathfrak{R}_{n}^{c}\left(k\right)=i\, n\cdot\tilde{A}_{b}\left(k\right)\delta\left(k\cdot n\right)\,2\mathrm{Tr}\left(t^{c}t^{b}\right)\\
+i^{2}g\int\frac{d^{4}p}{\left(2\pi\right)^{4}}\,\delta\left(k\cdot n\right)\, n\cdot\tilde{A}_{b}\left(p\right)\, n\cdot\tilde{A}_{b'}\left(k-p\right)\frac{i}{p\cdot n}\,2\mathrm{Tr}\left(t^{c}t^{b}t^{b'}\right)+\ldots\label{eq:[x]epslim}\end{multline}

From the above, we see that the operator $\mathfrak{R}_{n}\left(k\right)$
should be considered as a generalized function of $k\cdot n$. Therefore,
instead of the regularization (\ref{eq:[x]reg1}), (\ref{eq:[x]reg2})
one can use the following more practical prescription known from the
theory of the generalized functions; we use the infinite gauge links
(\ref{eq:gaugelink_[x]n}) with the following prescriptions for the
path-ordered integrals \begin{equation}
\frac{i}{p\cdot n+i\epsilon}=e^{isp\cdot n}\int_{-\infty}^{s}ds'\, e^{-is'p\cdot n}\label{eq:eikonal_prop}\end{equation}
and \begin{equation}
2\pi\delta\left(p\cdot n\right)=\int_{-\infty}^{\infty}ds\, e^{-is\, p\cdot n}.\label{eq:deltadef}\end{equation}
Note, that the $i\epsilon$ prescription in (\ref{eq:eikonal_prop})
is the same as used in \cite{Collins:1981uw,Collins:2011zzd} for
collinear PDFs. 

In the view of the above considerations, also the matrix element (\ref{eq:ME_def})
should be considered as a generalized function of $k\cdot n$. It
defines an object $\tilde{\mathcal{M}}_{n}$ we call \textit{off-shell
gauge invariant amplitude} for the process $g^{*}\left(k\right)g\left(k_{1}\right)\ldots g\left(k_{N}\right)\rightarrow0$,\begin{equation}
\mathfrak{M}\left(n,\varepsilon_{1},\ldots,\varepsilon_{N}\right)=\delta\left(k\cdot n\right)\delta^{4}\left(k_{A}+k_{1}+\ldots+k_{N}\right)\mathcal{\tilde{M}}_{n}\left(\varepsilon_{1},\ldots,\varepsilon_{N}\right).\label{eq:ME_deltas}\end{equation}
As a polarization vector for the off-shell gluon we understand a vector
$\varepsilon$ defined as \begin{equation}
\varepsilon_{\mu}=n^{\nu}D_{\mu\nu}\left(k\right),\end{equation}
where $D_{\mu\nu}\left(k\right)$ is the numerator of off-shell gluon
propagator. In particular in the Feynman gauge we have simply\begin{equation}
\varepsilon^{\mu}=n^{\mu}\end{equation}
and \begin{equation}
k\cdot\varepsilon=0.\end{equation}
If $n=p_{A}$, the amplitude $\tilde{\mathcal{M}}_{n}$ is the gauge
invariant high-energy off-shell amplitude from section \ref{sec:gauge_restoring_amp}. 

We present a simple and instructive analytic calculation for a process
$g^{*}gg$ in Appendix~\ref{sec:App_RGG}. It is of course very cumbersome
to do a similar calculation by hand for larger multiplicities. However,
the procedure is well defined and can be implemented in a computer
program, provided it can deal with many terms (see section \ref{sec:OGIME}).
Alternatively it is useful to construct the relevant Feynman rules.
Besides the standard QCD rules, the rules related to operator (\ref{eq:R_def})
insertion and eikonal propagators are needed. They will be constructed
in section \ref{sec:Feynman_rules}.

One can also define the matrix element of several gauge link operators,
with which one can define the gauge invariant amplitude $\mathcal{\tilde{M}}_{n_{A}n_{B}n_{C}\ldots}$
with several off-shell gluons with momenta $k_{A}$, $k_{B}$, $k_{C}$,
$\ldots$\begin{multline}
\mathfrak{M}\left(n_{A},n_{B},n_{C}\ldots,\varepsilon_{1},\ldots,\varepsilon_{N}\right)=\delta\left(k_{A}\cdot n_{A}\right)\delta\left(k_{B}\cdot n_{B}\right)\delta\left(k_{C}\cdot n_{C}\right)\ldots\\
\delta^{4}\left(k_{A}+k_{B}+k_{C}+\ldots+k_{1}+\ldots+k_{N}\right)\mathcal{\tilde{M}}_{n_{A}n_{B}n_{C}\ldots}\left(\varepsilon_{1},\ldots,\varepsilon_{N}\right)\\
\overset{*}{=}\left\langle k_{1},\varepsilon_{1},c_{1};\ldots;k_{N},\varepsilon_{N},c_{N}\right|\mathfrak{R}_{n_{A}}^{c_{A}}\left(k_{A}\right)\mathfrak{R}_{n_{B}}^{c_{B}}\left(k_{B}\right)\mathfrak{R}_{n_{C}}^{c_{C}}\left(k_{C}\right)\ldots\left|0\right\rangle \label{eq:ME_deltas-1-1-1}\end{multline}
by using the gauge links defined along $n_{A}$, $n_{B}$, $n_{C}$,$\ldots$.
Let us note that although in the present paper we limit ourselves
to gluonic on-shell states, the same prescription can be used for
any other on-shell state.

There is a limitation for the allowed vectors $n_{X}$, $X=A,B,\ldots$.
If some of $n_{X}$ are equal but not light-like, the parallel Wilson
lines start to interact. This causes a problem. This is most easily
seen when we realize that such interactions give rise to terms of
the form\begin{equation}
\frac{i}{k_{X_{1}}\cdot n_{X_{2}}}\,\, n_{X_{1}}\cdot n_{X_{2}},\,\,\, X_{1},X_{2}=A,B,C,\ldots\end{equation}
which is divergent for $n_{X_{1}}=n_{X_{2}}$, unless $n_{X_{1}}\cdot n_{X_{2}}=0$
(recall that $k_{X_{i}}\cdot n_{X_{i}}=0$, $i=1,2$). Therefore,
for definiteness, we assume that parallel Wilson lines have to be defined
using null vectors.

The construction (\ref{eq:ME_deltas-1-1-1}) can be used in the following
way. Suppose we have a Green's function, which is reduced in such
a way that some of the legs are on-shell and have standard polarization
vectors, while some remain off-shell with momenta $k_{A}$, $k_{B}$,
$\ldots$ and are contracted with certain {}``polarization vectors''
$\varepsilon_{A}$,$\varepsilon_{B}$,$\ldots$. Call this reduced
Green's function an off-shell amplitude $\mathcal{M}_{\varepsilon_{A}\varepsilon_{B\ldots}}$
--- it is not gauge invariant. In order to find a gauge invariant extension
$\tilde{\mathcal{M}}_{n_{A}n_{B\ldots}}$, we choose the vectors $n_{A}$,
$n_{B}$, $\ldots$ satisfying \begin{equation}
k_{X}\cdot n_{X}=0,\,\,\,\varepsilon_{X\,\mu}=n_{X}^{\nu}D_{\mu\nu}\left(k_{X}\right),\,\, X=A,B,\ldots\end{equation}
and use (\ref{eq:ME_deltas-1-1-1}) with the gauge links directions
along $n_{A}$,$n_{B}$,$\ldots$. In the Feynman gauge the gauge
links directions correspond to polarization vectors of the off-shell
gluons \begin{equation}
\varepsilon_{X}=n_{X},\,\, X=A,B,\ldots.\end{equation}
The contribution to (\ref{eq:ME_deltas-1-1-1}) coming from the first
nontrivial term in expansion of the gauge links is precisely the starting
non-gauge-invariant amplitude $\mathcal{M}_{n_{A}n_{B\ldots}}$ while
the rest form an analog of $\mathcal{W}$ from section \ref{sec:gauge_restoring_amp}.

Since the delta functions $\delta\left(k_{X}\cdot n_{X}\right)$ in
(\ref{eq:ME_deltas-1-1-1}) come entirely from the insertion of $\mathfrak{R}_{n_{X}}^{c_{X}}\left(k_{X}\right)$
operators, one can define a more physical quantity by integrating
the operators over the arguments of the deltas. To this end let us
decompose any four vector to the component along $n_{X}$ as follows
\begin{equation}
v^{\mu}=v^{\left(n_{X}\right)}n_{X}^{\mu}+v^{\left(n\right)}n^{\mu}+v_{T}^{\left(n_{X},n\right)\mu}\label{eq:decomp}\end{equation}
where $v_{T}^{\left(n_{X},n\right)}\cdot n_{X}=v_{T}^{\left(n_{X},n\right)}\cdot n=0$
and $n$ is an arbitrary four vector such that $n_{X}\cdot n\neq0$.
Using this we define \begin{equation}
\int dk_{X}^{\left(n_{X}\right)}\mathfrak{R}_{n_{X}}^{c_{X}}\left(k_{X}\right)\equiv\mathcal{R}_{n_{X}}^{c_{X}}\left(k_{X}^{\left(n\right)},k_{T\, X}^{\left(n_{X}\right)}\right).\label{eq:Rphys_def}\end{equation}
Note, that the gauge link defining $\mathcal{R}$ does not depend
on the $x$ component along $n$; the integration over $k_{X}^{\left(n_{X}\right)}$
gives the delta function $\delta\left(x^{\left(n_{X}\right)}n_{X}^{2}+x^{\left(n\right)}n_{X}\cdot n\right)$
(c.f. the definition (\ref{eq:R_def})) which in turn can be integrated
over $x^{\left(n\right)}$ residing inside $d^{4}x$ measure in (\ref{eq:R_def}). 

The action of the operator (\ref{eq:Rphys_def}) on the vacuum state
may be considered as a creation of a certain state corresponding to
an off-shell gluon (in the terminology of \cite{Antonov:2004hh}
they would correspond to reggeized gluons if $n_{X}$ is eikonal momentum)\begin{equation}
\left\langle k_{X},n_{X},c_{X}\right|=\left\langle 0\right|\mathcal{R}_{n_{X}}^{c_{X}}\left(k_{X}^{\left(n\right)},k_{T\, X}^{\left(n_{X}\right)}\right).\label{eq:Rstate}\end{equation}
Such a state belongs to the cohomology of the Becchi-Rouet-Stora-Tyutin
(BRST) transformation, similar to other {}``physical'' asymptotic
states as gluons or quarks. Although in principle it follows from
the gauge invariance of infinite gauge links, we check this fact explicitly
in Appendix \ref{sec:BRST_R}, as it may not be clear that the prescription
for path-ordered integrals preserves this property (see also section
\ref{sec:Word_ident}).

\section{Automatic calculation of matrix elements}

\label{sec:OGIME}

Having the definition (\ref{eq:ME_deltas-1-1-1}) together with the
prescriptions (\ref{eq:eikonal_prop})-(\ref{eq:deltadef}) the calculation
of any tree level amplitude is a purely algebraic task. The only integrals
to be performed are of exponential nature, which in turn can also
be done symbolically. Therefore, we have constructed a program written
in FORM \cite{Kuipers:2012rf}, that calculates (\ref{eq:ME_deltas-1-1-1})
automatically. It does not refer to the Feynman rules, instead it
uses the Wick theorem (see an example calculation presented in the
Appendix \ref{sec:App_RGG}). Therefore in the future fermionic and
other gauge fields can be added relatively easy. Although FORM is
very powerful and can deal with huge expressions, the number of terms
that appear due to the Wick theorem is often enormous. Therefore it
was a crucial task to find a reasonable algorithm to deal with the
Wick contractions. At the moment the program was tested on amplitudes
with time ordered product of at most 13 gauge fields on a standard
laptop. This corresponds for example to a matrix element of four Wilson
line operators and a gluon. Another difficulty to overcome was related
to a simplification of expressions; it is necessary to use momentum
conservation and relations between invariants at the intermediate
steps of calculation and the program does it automatically. The color
algebra is also done automatically, so that the results are given
in the color-ordered representation \cite{Mangano:1987xk,Mangano:1990by}.

The program is called $\mathtt{OGIME}$ --- an alias for Off-shell
Gauge Invariant Matrix Elements. It is available from the author's
web pages \cite{Kotko_OGIME} or via email upon request. The technical
details concerning the program are beyond the scope of this paper
and will be presented elsewhere.

\section{The Feynman rules}

\label{sec:Feynman_rules}

In order to construct the relevant Feynman rules for an insertion
of $\mathfrak{R}_{n_{X}}^{c_{X}}\left(k_{X}\right)$ operators in
the QCD matrix element, let us again consider one of the terms in
its expansion, say the $m$-th term \begin{multline*}
\frac{1}{2\pi g}\left(ig\right)^{m}\int d^{4}xe^{-ik_{X}\cdot x}\int_{-\infty}^{\infty}ds_{1}\int_{-\infty}^{s_{1}}ds_{2}\ldots\int_{-\infty}^{s_{m-1}}ds_{m}\\
\, n_{X}\cdot A_{b_{1}}\left(x+s_{1}n_{X}\right)\ldots n_{X}\cdot A_{b_{m}}\left(x+s_{m}n_{X}\right)\,2\mathrm{Tr}\left(t^{c_{X}}t^{b_{1}}\ldots t^{b_{m}}\right)\\
=\int\frac{d^{4}p_{1}}{\left(2\pi\right)^{4}}\ldots\frac{d^{4}p_{m}}{\left(2\pi\right)^{4}}\,\delta^{4}\left(p_{1}+\ldots+p_{m}+k_{X}\right)\int_{-\infty}^{\infty}\frac{ds_{1}}{\pi g}\ldots\int_{-\infty}^{s_{m-1}}ds_{m}\\
\delta_{B_{m}A_{0}}t_{A_{0}B_{0}}^{c_{X}}e^{-is_{1}p_{1}\cdot n_{X}}\,\delta_{B_{1}A_{2}}e^{-is_{2}p_{2}\cdot n_{X}}\ldots\delta_{B_{m-1}A_{m}}e^{-is_{m}p_{m}\cdot n_{X}}\,\\
igt_{A_{1}B_{1}}^{b_{1}}n_{X}\cdot\tilde{A}_{b_{1}}\left(p_{1}\right)\ldots igt_{A_{m}B_{m}}^{b_{m}}n_{X}\cdot\tilde{A}_{b_{m}}\left(p_{m}\right),\end{multline*}
where $\tilde{A}$ are the Fourier-transformed gluon fields and the
capital letter indices $A_{i}$, $B_{i}$ are the fundamental color
indices (summed over). The above structure can be interpreted as follows.
The fields are separated on a line, graphically represented by a double
line carrying color quantum numbers in fundamental representation
and the total momentum $k_{X}$ which we assign to its beginning.
Each gluon field with the color $b$ is attached to this line via
$ig\, t_{AB}^{b}\, n_{X}^{\mu}$ vertex. The field attachments are
separated by eikonal propagators which follow from the integrals along
the path:\begin{multline}
\int_{-\infty}^{\infty}\frac{ds_{1}}{\pi g}\ldots\int_{-\infty}^{s_{m-1}}ds_{m}\, t_{A_{0}B_{0}}^{c_{X}}\delta_{B_{m}A_{0}}e^{-is_{1}p_{1}\cdot n_{X}}\,\delta_{B_{1}A_{2}}e^{-is_{2}p_{2}\cdot n_{X}}\ldots\delta_{B_{m-1}A_{m}}e^{-is_{m}p_{m}\cdot n_{X}}\\
=\int_{-\infty}^{\infty}\frac{ds_{1}}{\pi g}\ldots\int_{-\infty}^{s_{m-2}}ds_{m-1}\,\delta_{B_{m}A_{0}}t_{A_{0}B_{0}}^{c_{X}}\\
e^{-is_{1}p_{1}\cdot n_{X}}\ldots\delta_{B_{m-2}A_{m-1}}e^{-is_{m-1}\left(p_{m-1}+p_{m}\right)\cdot n_{X}}\frac{i\delta_{B_{m-1}A_{m}}}{p_{m}\cdot n_{X}+i\epsilon}\\
=\frac{2\delta_{B_{m}A_{0}}t_{A_{0}B_{0}}^{c_{X}}}{g}\,\delta\left(p_{1}\cdot n_{X}+\ldots p_{m}\cdot n_{X}\right)\,\frac{i\delta_{B_{1}A_{2}}}{\left(p_{1}+\ldots+p_{m}\right)\cdot n_{X}+i\epsilon}\ldots\frac{i\delta_{B_{m-1}A_{m}}}{p_{m}\cdot n_{X}+i\epsilon}.\label{eq:FR_1}\end{multline}
Above, we have used $i\epsilon$ prescription (\ref{eq:eikonal_prop})
as explained in the previous section. In the last line we encounter
the delta function (\ref{eq:deltadef}), which together with momentum
conservation will give the delta function $\delta\left(k_{X}\cdot n_{X}\right)$.
The factor 2 in front of the color delta function will assure the
correct matching with the standard QCD result when the on-shell limit
is taken (of course after multiplying by inverse propagators for $k_{X}$
momenta).

\begin{figure}
\begin{centering}
\includegraphics[width=0.4\textwidth]{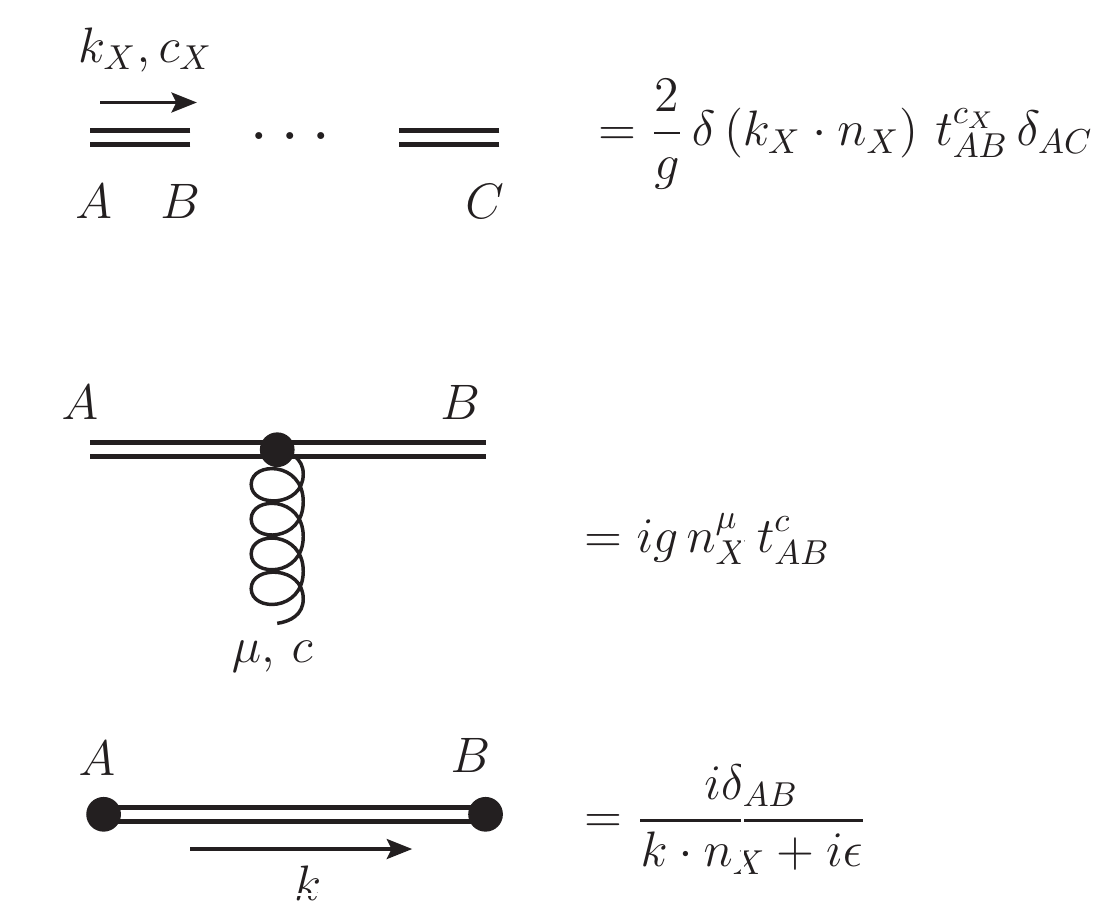}
\par\end{centering}

\caption{\small The Feynman rules for an insertion of the operator $\mathfrak{R}_{n_{X}}^{c_{X}}\left(k_{X}\right)$.
The top diagram is a {}``skeleton'' for a gauge link; it consist
in the transversality constraint, the color projection, and color
delta function eventually giving a trace. The right-most end of the
double line has zero momentum. The middle diagram represents the coupling
of a gluon to the gauge link. The bottom diagram represents the gauge-link
propagator originating in path-ordered integrals. \label{fig:gaugelink_FR2}}
\end{figure}

In figure~\ref{fig:gaugelink_FR2} we have gathered the resulting Feynman
rules for insertion of (\ref{eq:R_def}) into a QCD matrix element.
The top diagram is a {}``skeleton'' for a gauge link and originates
in the delta function in (\ref{eq:FR_1}) with the prefactor. Its
left-most side has the momentum and adjoint color quantum number of
the $\mathfrak{R}_{n_{X}}^{c_{X}}\left(k_{X}\right)$ operator. The
momentum (and the color) flows to the right. The end of the gauge
link (the right-most part) has momentum zero. Between the beginning
and the end of the Wilson line gluons can be attached, according to
the rest of the rules given in figure~\ref{fig:gaugelink_FR2}.

\begin{figure}
\begin{centering}
\includegraphics[width=0.4\textwidth]{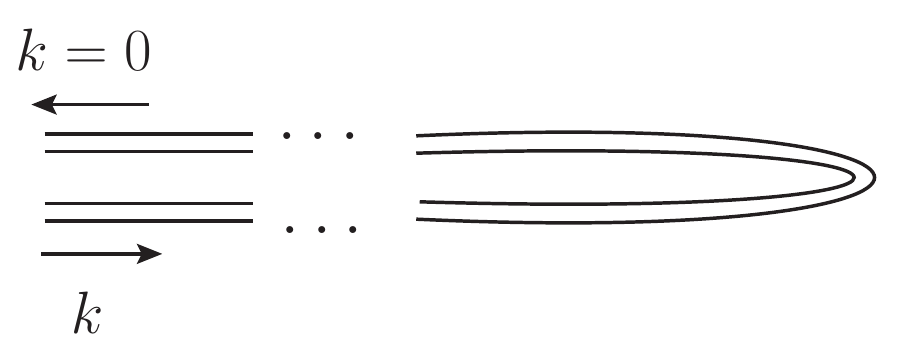}
\par\end{centering}

\caption{\small The improved Feynman rule for the gauge link (the dots stand
for gluon attachments and gauge-link propagators). Such a rule is
more convenient when more gauge links are present and is easy to convert
to color-ordered rule. \label{fig:bent_gaugelink}}
\end{figure}

In order to better trace the color (and momentum) flow in a diagram
(especially when more gauge links are present) it is convenient to
{}``bend'' the gauge link as shown in figure~\ref{fig:bent_gaugelink}.
In case of two off-shell gluons, one can alternatively draw the two
gauge links as the top-most and the most bottom features, remembering
that the direction of a momentum flow (and the trace direction) has
to be reversed for the bottom-most gauge link.

In the following it will be convenient to work with color-ordered
amplitudes defined in eq.~(\ref{eq:Color_ord_def}). The standard
color-ordered QCD Feynman rules are listed e.g.\ in the appendix
of \cite{Mangano:1990by} or \cite{vanHameren:2012uj}. We have
to supply this list by the rules for gauge links. This is straightforward,
as gauge link contribution gives always a color trace, similar to
a gluon sub-amplitude (recall that gluon vertices give products of
traces due to $f_{abc}=-2i\left(\mathrm{Tr}\left(t^{a}t^{b}t^{c}\right)-\mathrm{Tr}\left(t^{c}t^{b}t^{a}\right)\right)$).
We omit color indices, drop all the color matrices and assign $1/\sqrt{2}$
for each coupling of a gluon to a gauge link. Further, we consider
only planar diagrams with the {}``bent'' gauge link (of course we
can {}``unroll'' it keeping track of the color and momentum flow).
Note, however that our definition (\ref{eq:Color_ord_def}) assumes
the standard normalization of the color generators $\mathrm{Tr}\left(t^{a}t^{b}\right)=\delta^{ab}/2$,
while the usual color-ordered Feynman rules do assume that $\mathrm{Tr}\left(t^{a}t^{b}\right)=\delta^{ab}$
(see e.g.\ \cite{Mangano:1990by}). In order to compensate for this
mismatch one has to multiply the result by a factor $\left(\sqrt{2}\right)^{N}$,
where $N$ is the number of external on-shell momenta.

\section{The Ward identities}

\label{sec:Word_ident}

We have seen in the beginning of the section \ref{sec:formal_develop}
that the gauge links defining the operator (\ref{eq:R_def}) are manifestly
gauge invariant with respect to small gauge transformations. However,
the path integrals residing inside the expansion are divergent and
a regularization is required (or a prescription to give a physical
meaning to the integrals). The regularization leads to finite-length
Wilson lines, thus violating the gauge invariance argument. After
the limit $\epsilon\rightarrow0$ is taken the gauge invariance should
be recovered. Therefore it is necessary and instructive to prove that
the off-shell amplitudes defined via (\ref{eq:ME_deltas-1-1-1}) together
with the prescription to define the path integrals (\ref{eq:eikonal_prop})
do satisfy the Ward identities. Moreover, this exercise nicely illustrates
the reason the method of ref.~\cite{vanHameren:2012uj} (and recalled
in section \ref{sec:gauge_restoring_amp}) does actually work. We
consider only gluonic amplitudes hereafter; the amplitudes with quarks
can be analyzed even more easily in a similar manner. For another
study of the Ward identities in the context of Lipatov's effective
action see \cite{Bartels:2012mq}.

\subsection{Preliminaries}

\label{sub:WardId_prelim}

Before we face our main task, we will gather some preliminary results
that will be useful later. Let us start with the Slavnov-Taylor identity
for partially reduced Green's function. Suppose we have the Green's
function (in the Feynman gauge) and we contract one of the external
legs with the corresponding momentum. The Slavnov-Taylor identity
states that it equals to the sum of contributions with ghosts, diagrammatically

\noindent \begin{tabular}{>{\centering}m{0.9\columnwidth}>{\centering}m{0.05\columnwidth}}
\bigskip{}

\centering{}\includegraphics[height=0.058\paperheight]{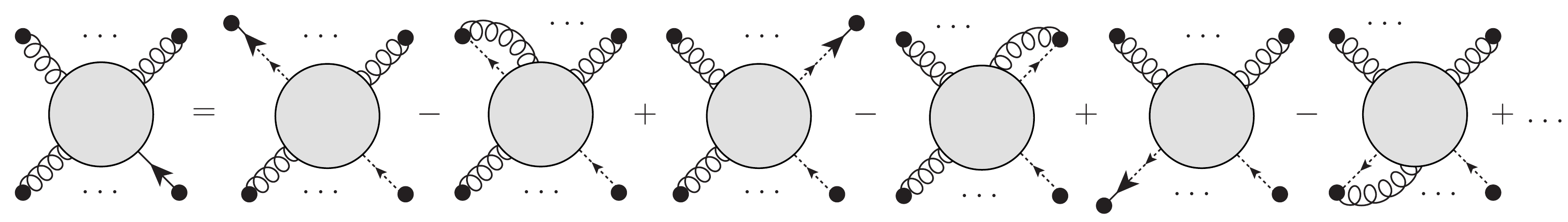} & \centering{}
(\myref)\label{ST_basic_1}
\tabularnewline
\end{tabular}

\noindent The large arrows denote the contraction with the momentum
of the line, the big dots indicate that the line has a propagator,
whereas the horizontal dots between the upper and bottom legs remind
that there are several legs of the same type. The dot with a ghost
and a gluon connected to it is the the vertex due to the BRST transformation,
namely $gf^{abd}A_{\mu}^{b}c^{d}$ with $c^{d}$ being a ghost field.
Finally, the ellipses after the plus sign denote similar diagrams
for the remaining legs. Suppose now that we want to reduce the Green's
function in such a way that the bottom legs are on-shell. The standard
reduction formula is applied to those lines, i.e. they are multiplied
by the inverse propagators and contracted with the polarization vectors.
After such reduction, we are left only with the diagrams where the
outgoing ghosts are off-shell (they exit in the upper part of the
diagrams from \eqref{ST_basic_1}).

\begin{tabular}{>{\centering}m{0.83\columnwidth}>{\centering}m{0.05\columnwidth}}
\bigskip{}

\centering{}\includegraphics[height=0.058\paperheight]{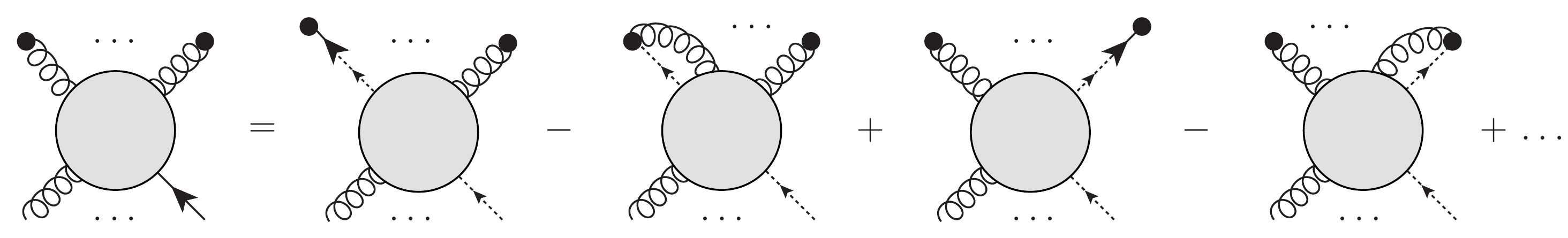} & \centering{}
(\myref)\label{ST_basic_2}
\tabularnewline
\end{tabular}

\noindent Above we have indicated that the bottom legs are on-shell
(the big dots are missing there because the propagators were amputated)
while the top are still off-shell.

In what follows, in order to simplify the diagrammatic analysis we
shall work with color ordered amplitudes. Moreover, for further reduction
of the number of diagrams we shall work in the axial gauge for internal
propagators with the gauge vector $n'$, in general not light-like.
For that choice of gauge the identity \eqref{ST_basic_2} can still
be applied (c.f. \cite{vanHameren:2012uj}); in the axial gauge ghosts
couple to gluons via $gf^{abc}n'^{\mu}$ and the ghost propagator
is $i/\left(k\cdot n'+i\epsilon\right)$ .

Consider now an attachment of a sub-amplitude with $N$ external on-shell
gluons to the gauge link defined via vector $n$, $n\neq n'$. When
we replace one of the polarization vectors by the corresponding momentum
of the line we get

\begin{tabular}{>{\centering}m{0.83\columnwidth}>{\centering}m{0.05\columnwidth}}
\bigskip{}

\centering{}\includegraphics[height=0.078\paperheight]{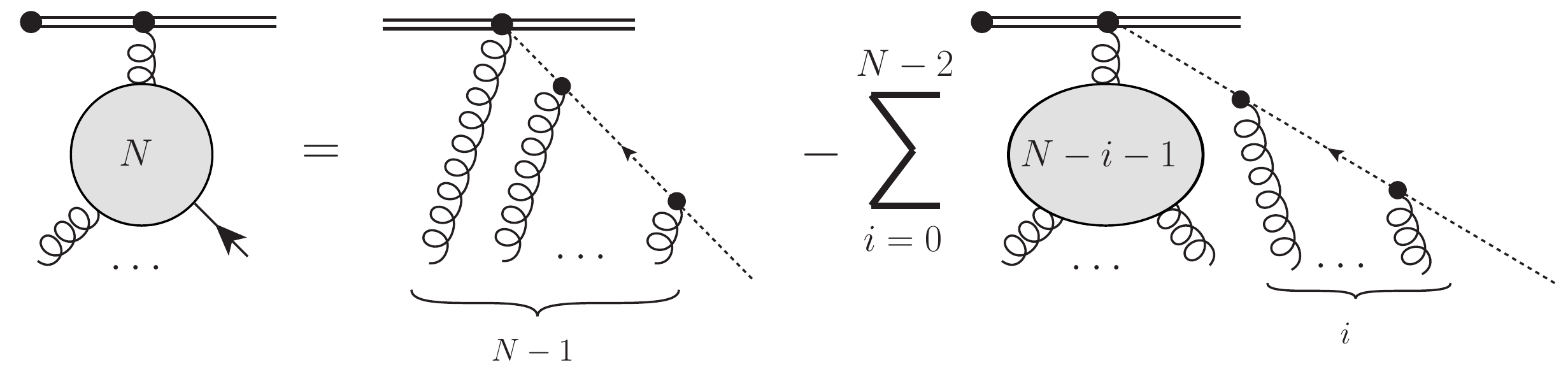} & \centering{}
(\myref)\label{ST_blob}
\tabularnewline
\end{tabular}

\noindent where the numbers on the blobs denote the number of external
on-shell legs. This identity follows directly from the Slavnov-Taylor
identity \eqref{ST_basic_2} in the axial gauge and the transversality
of the axial-gauge gluon propagator to the gluon-ghost coupling, i.e. 

\begin{tabular}{>{\centering}m{0.83\columnwidth}>{\centering}m{0.05\columnwidth}}
\bigskip{}

\centering{}\includegraphics[height=0.056\paperheight]{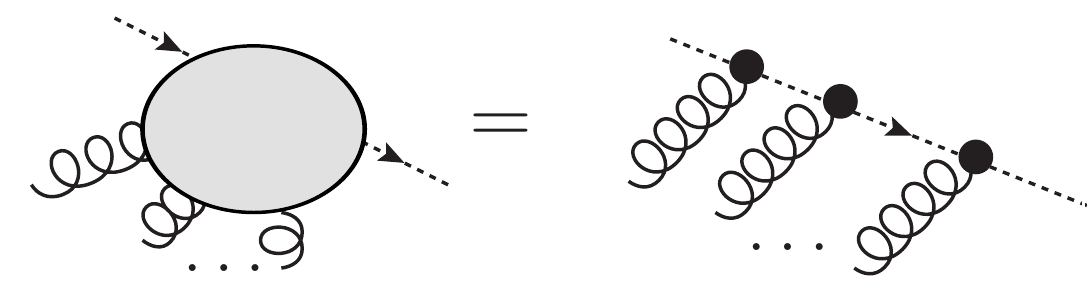} & \centering{}
(\myref)\label{ST_axial}
\tabularnewline
\end{tabular}

\noindent Note, that the first term on the r.h.s of \eqref{ST_blob}
has no gauge-link propagator as it was canceled by the outgoing momentum
of the ghost contracted with $n$

\begin{tabular}{>{\centering}m{0.83\columnwidth}>{\centering}m{0.05\columnwidth}}
\bigskip{}

\centering{}\includegraphics[height=0.035\paperheight]{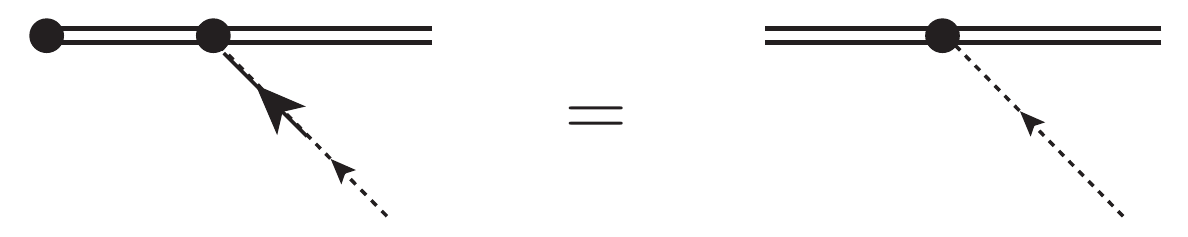} & \centering{}
(\myref)\label{ST_prop}
\tabularnewline
\end{tabular}

\noindent The color-ordered rule for the BRST gluon-ghosts coupling
is taken with the plus sign if the ghost is to the right of the gluon
(an exchange leads to a minus sign). We use here the {}``unbent''
gauge link with momentum flowing from the left to the right.

In order to derive the Ward identity for a gauge link, consider now
the set of all the contributions attached to a gauge link, starting
from a certain point (i.e. we cut the gauge link at some point including
a propagator)

\begin{tabular}{>{\centering}m{0.83\columnwidth}>{\centering}m{0.05\columnwidth}}
\bigskip{}

\raggedright{}\includegraphics[height=0.06\paperheight]{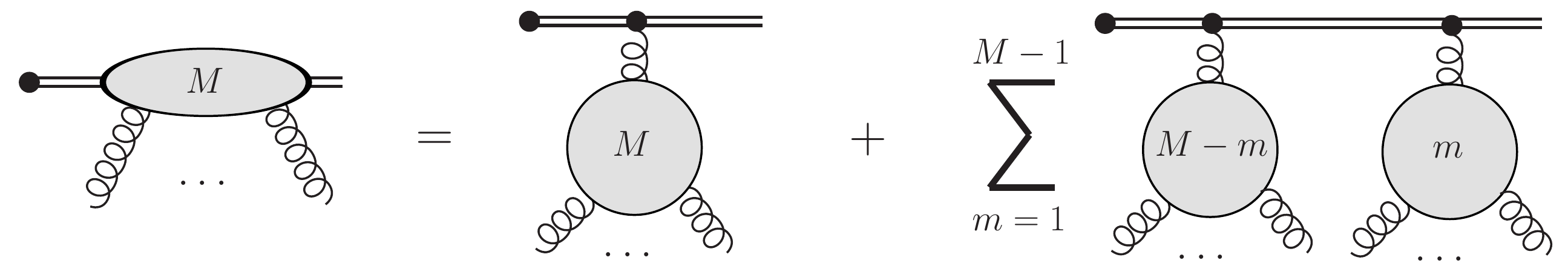} & \centering{}\tabularnewline
\raggedleft{}\includegraphics[height=0.06\paperheight]{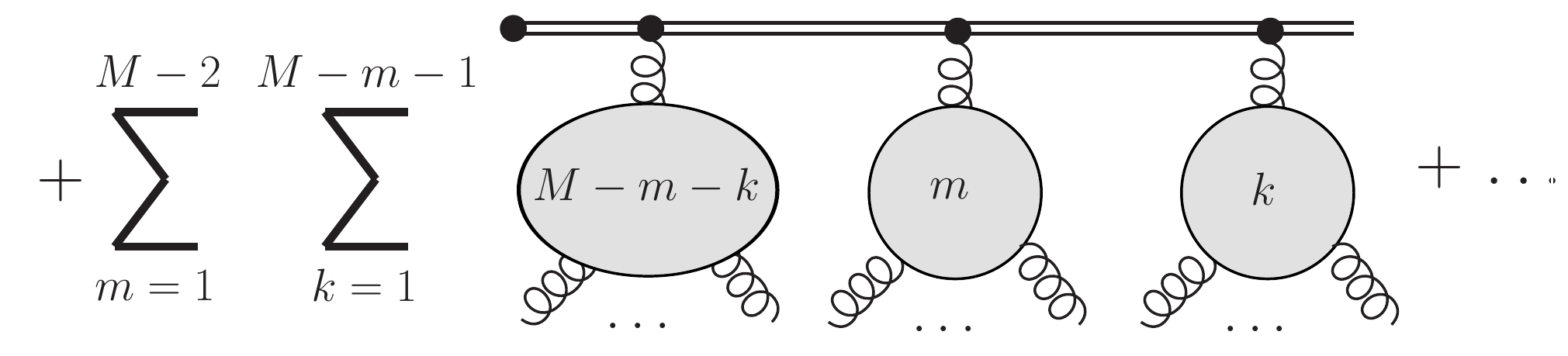} & \centering{}
(\myref)\label{Blob_expansion}
\tabularnewline
\end{tabular}

\noindent where the ellipses after the plus sign denote the contributions
with more blobs attached to the gauge link. Let us now replace the
polarization vector by the momentum for the rightmost gluon. Since
for two consecutive blobs we have (following \eqref{ST_blob})

\begin{tabular}{>{\centering}m{0.83\columnwidth}>{\centering}m{0.05\columnwidth}}
\bigskip{}

\raggedright{}\includegraphics[height=0.07\paperheight]{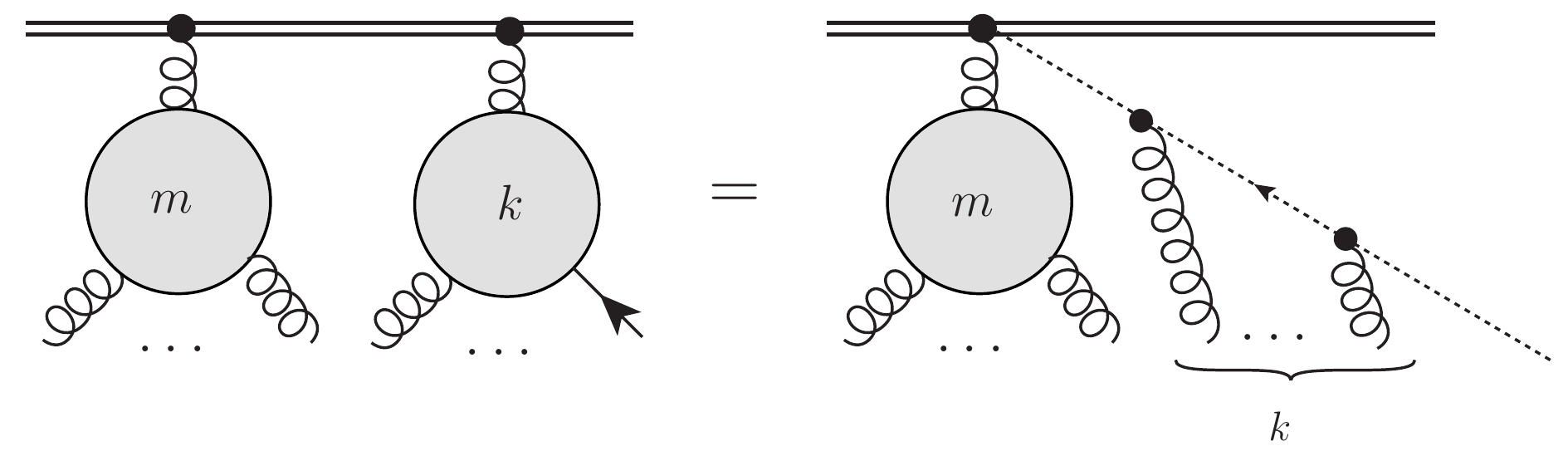} & \centering{}\tabularnewline
\raggedleft{}\includegraphics[height=0.07\paperheight]{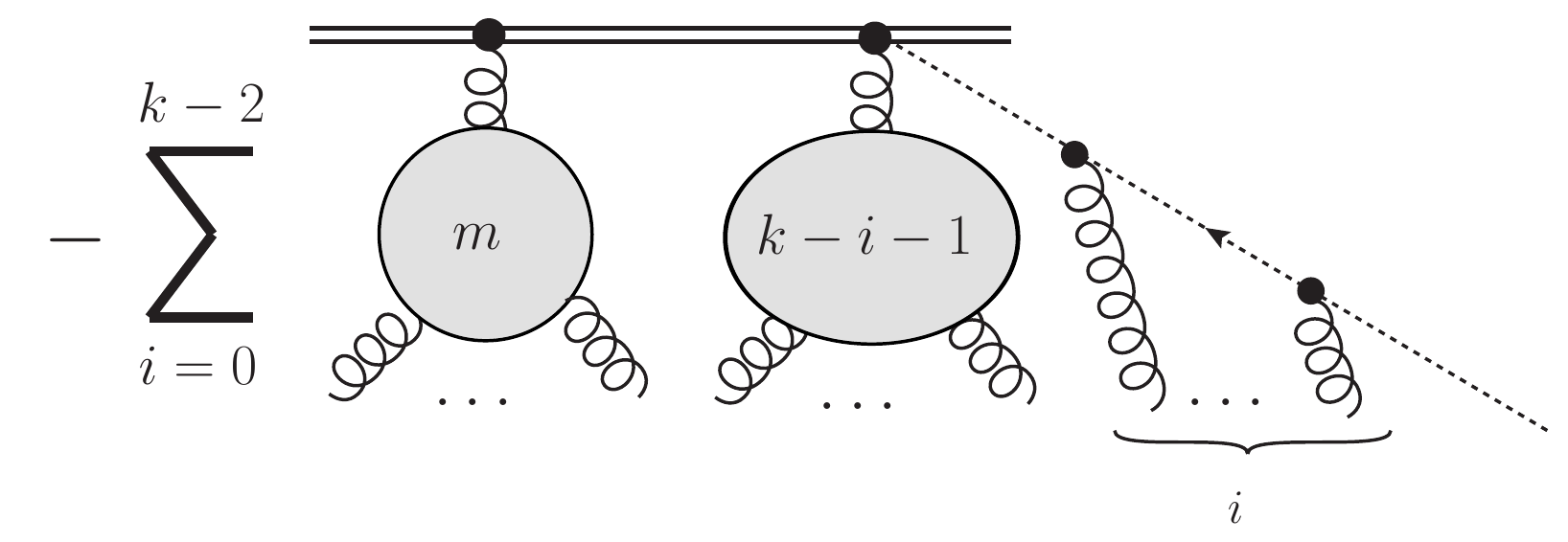} & \centering{}
(\myref)
\tabularnewline
\end{tabular}

\noindent we find that the Ward identity for the whole series of
the blobs attached to the gauge link takes the following simple form

\begin{tabular}{>{\centering}m{0.83\columnwidth}>{\centering}m{0.05\columnwidth}}
\bigskip{}

\centering{}\includegraphics[height=0.08\paperheight]{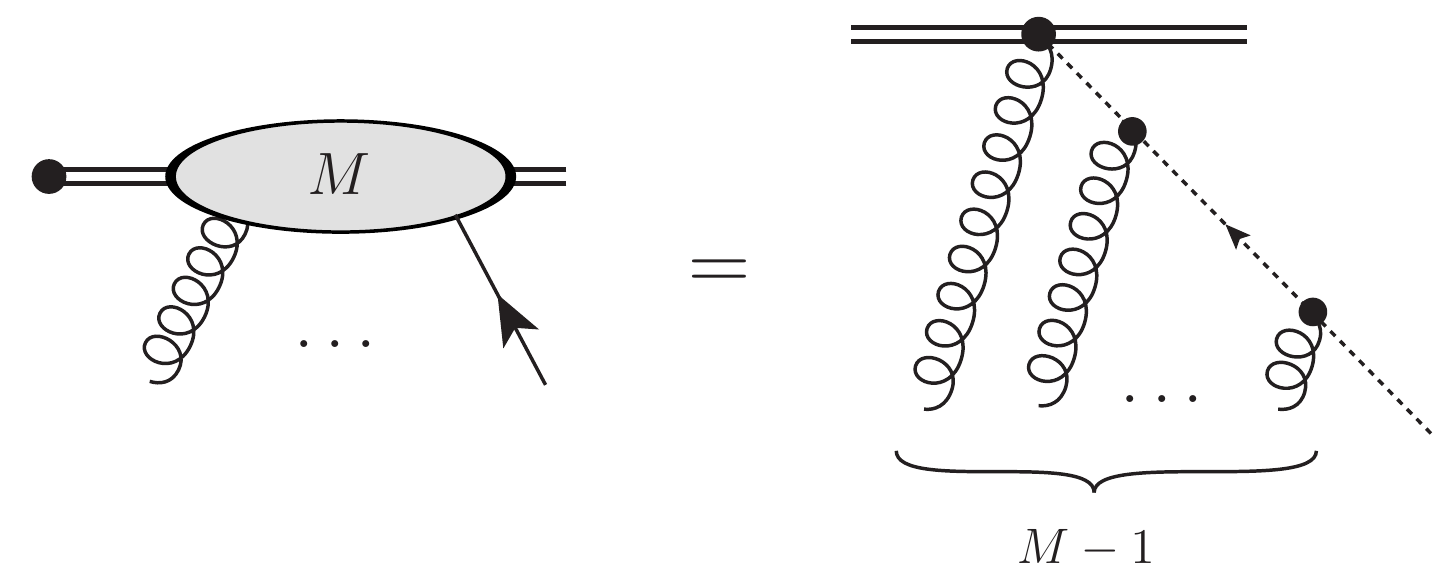} & \centering{}
(\myref)\label{Ward_gl}
\tabularnewline
\end{tabular}

\noindent All the terms canceled between the consecutive blobs in
\eqref{Blob_expansion}, except the first term.

Similar identities can be derived (with somewhat more effort) for
the replacements of the other gluons then the rightmost. Let us now
discuss the Ward identities for (\ref{eq:ME_deltas-1-1-1}) for one
and two Wilson line operator insertions.

\subsection{One off-shell gluon}

Since for a single off-shell gluon we have one Wilson line operator,
the Ward identity follows directly from \eqref{Ward_gl}. Let us
suppose the gauge link is defined with the momentum $k_{A}$ and the
direction $n_{A}$. We have

\begin{tabular}{>{\centering}m{0.83\columnwidth}>{\centering}m{0.05\columnwidth}}
\bigskip{}

\centering{}\includegraphics[height=0.08\paperheight]{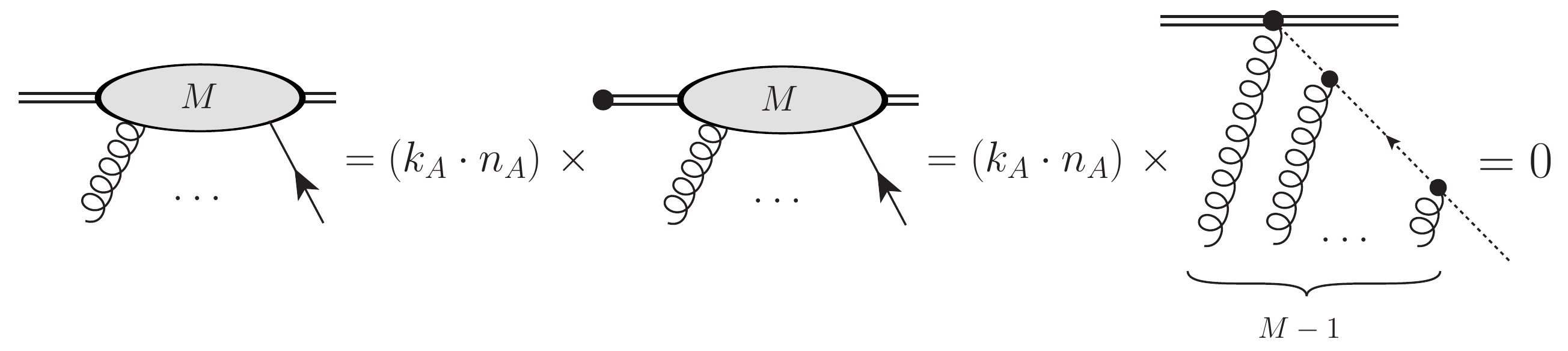} & \centering{}
(\myref)\label{Ward_R}
\tabularnewline
\end{tabular}

\noindent To obtain the first equality we have multiplied and divided
by the gauge link propagator. The second equality follows directly
from \eqref{Ward_gl}. Finally, the last equality follows in the
distributional sense, i.e. after integration over $k^{\left(n_{A}\right)}$
due to the delta $\delta\left(k_{A}\cdot n_{A}\right)$ which resides
in the gauge link.

Let us note, that when the axial gauge vector $n'=n_{A}$, the only
contributions to \eqref{ST_basic_2} are precisely of the form
of the rightmost term of \eqref{Ward_R}. This leads to the conclusion,
that in such a case the gauge link can be mimicked by the sum of the
Slavnov-Taylor gauge terms, with the external ghosts traded to longitudinal
gluons projected on $n_{A}$. This is precisely what has been used
in ref.~\cite{vanHameren:2012uj}.

\subsection{Two off-shell gluons}

Now let us consider the situation with two Wilson line operators.
Let us choose the momentum and the direction of the first gauge link
to be $k_{A}$ and $n_{A}$, while the for the second we choose $k_{B}$
and $n_{B}$. The relevant matrix element is graphically expressed
as

\begin{tabular}{>{\centering}m{0.83\columnwidth}>{\centering}m{0.05\columnwidth}}
\bigskip{}

\centering{}\includegraphics[height=0.076\paperheight]{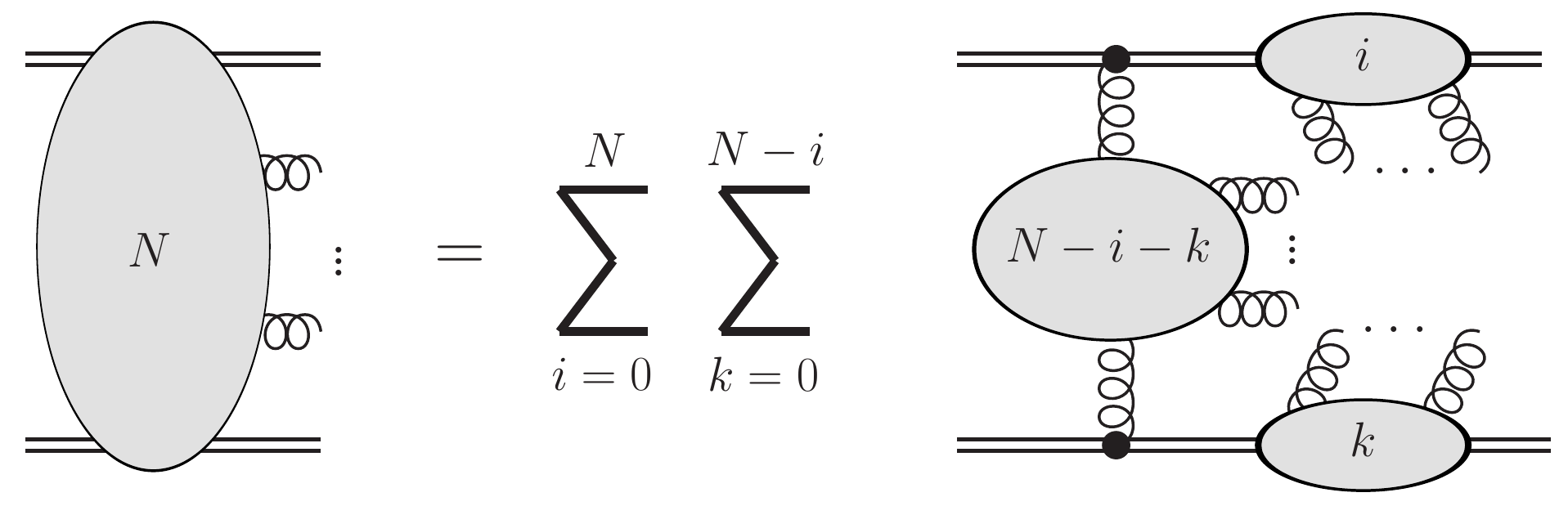} & \centering{}
(\myref)\label{Ward_RR_1}
\tabularnewline
\end{tabular}

\noindent Since we use here the {}``unbent'' Feynman rule for the
gauge links (see the penultimate paragraph of section \ref{sec:Feynman_rules}),
one has to remember that for the bottom gauge link the momentum flows
from the right to the left (the opposite to the top one). Let us now
replace the polarization vector of the top on-shell gluon by its momentum.
It is convenient to split \eqref{Ward_RR_1} into three distinct
topologies, i.e. we consider

\begin{flushleft}
\begin{tabular}{>{\centering}m{0.9\columnwidth}>{\centering}m{0.05\columnwidth}}
\bigskip{}

\raggedright{}\includegraphics[height=0.09\paperheight]{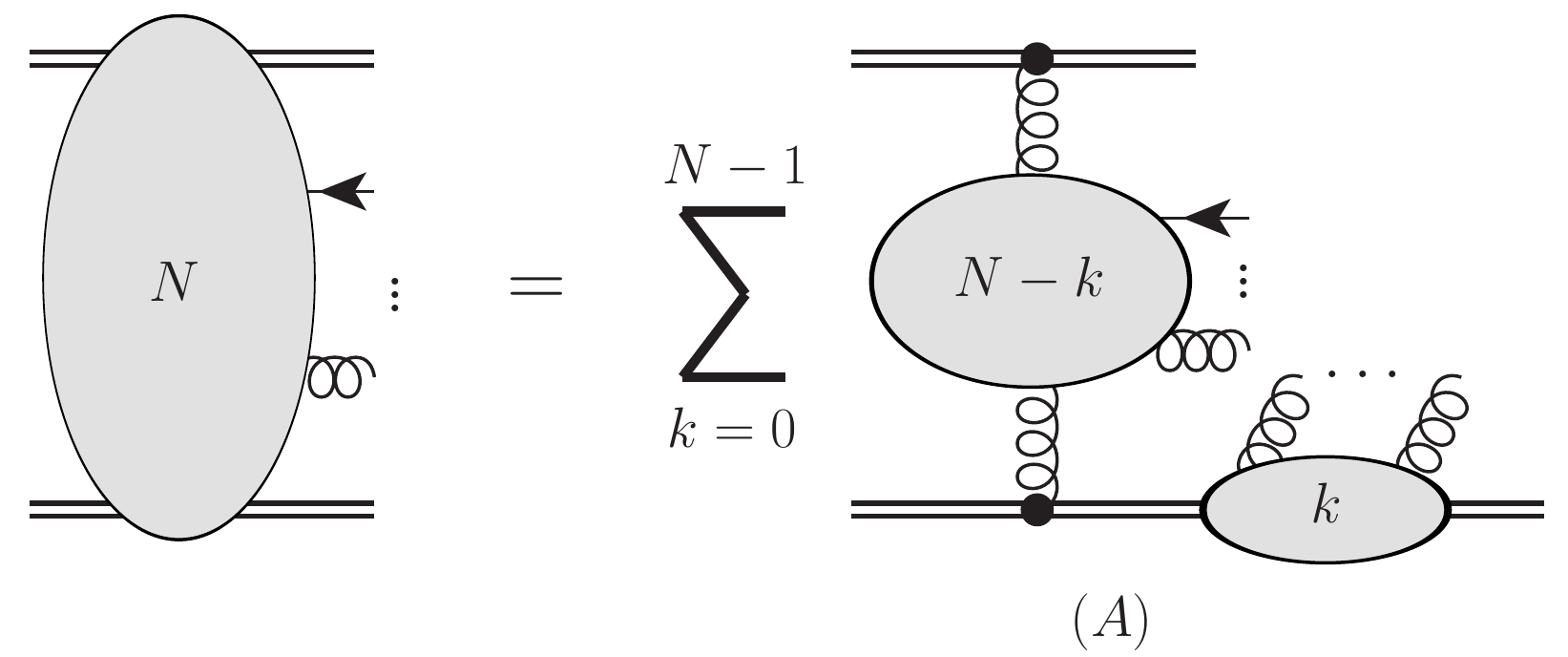} & \centering{}\tabularnewline
\end{tabular}
\par\end{flushleft}

\begin{flushleft}
\begin{tabular}{>{\centering}m{0.9\columnwidth}>{\centering}m{0.05\columnwidth}}
\raggedleft{}\includegraphics[height=0.09\paperheight]{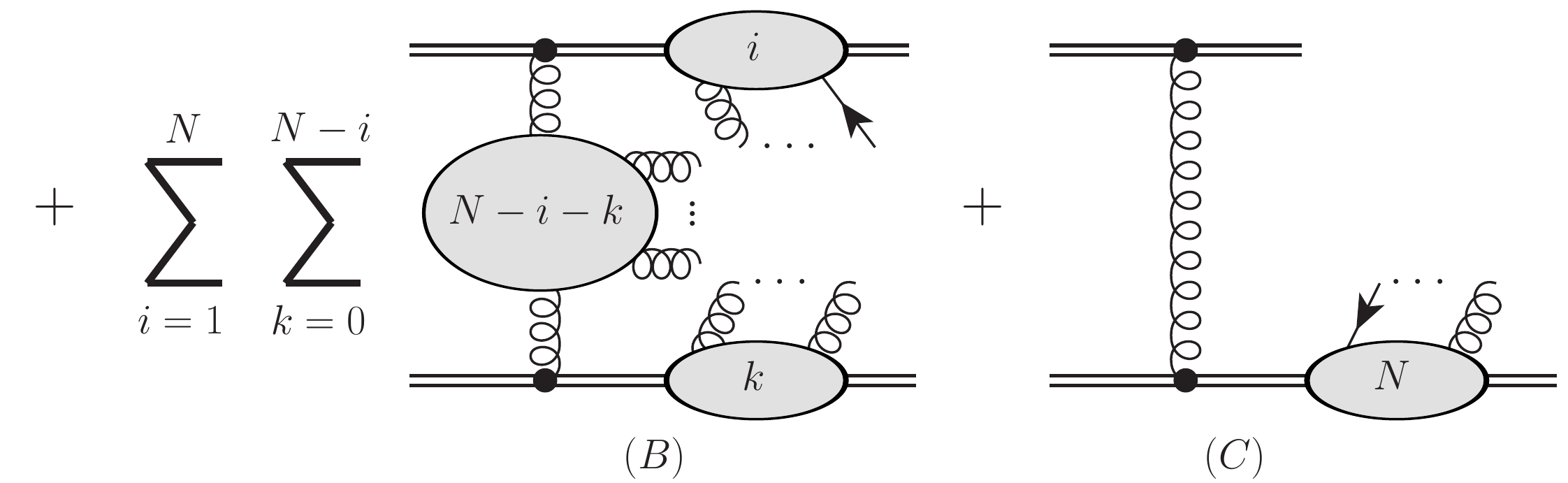} & \centering{}
(\myref)\label{Ward_RR_2}
\tabularnewline
\end{tabular}
\par\end{flushleft}

\noindent Consider now the $\left(A\right)$ and $\left(B\right)$
terms. Using \eqref{Ward_gl}, \eqref{ST_basic_2} and the fact
that $k_{A}\cdot n_{A}=0$ due to the delta function $\delta\left(k_{A}\cdot n_{A}\right)$
residing in the upper gauge link, they become

\begin{flushleft}
\begin{tabular}{>{\centering}m{0.83\columnwidth}>{\centering}m{0.05\columnwidth}}
\bigskip{}

\raggedright{}\includegraphics[height=0.09\paperheight]{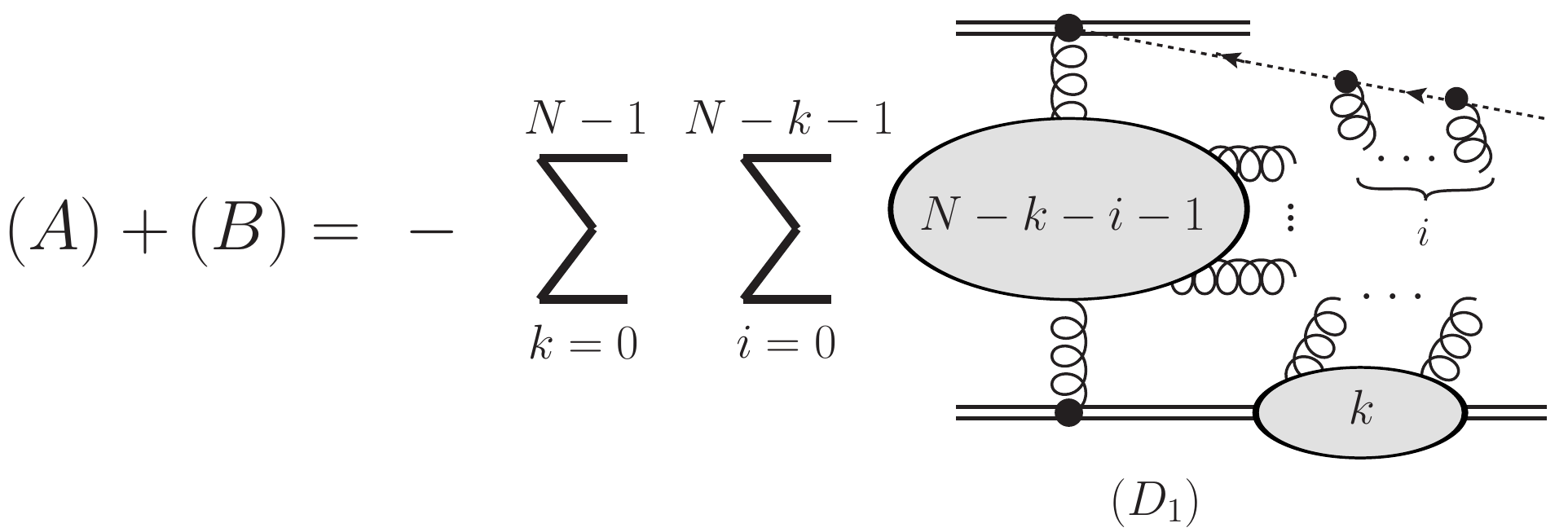} & \centering{}\tabularnewline
\end{tabular}
\par\end{flushleft}

\begin{flushleft}
\begin{tabular}{>{\centering}m{0.87\columnwidth}>{\centering}m{0.05\columnwidth}}
\raggedleft{}\includegraphics[height=0.09\paperheight]{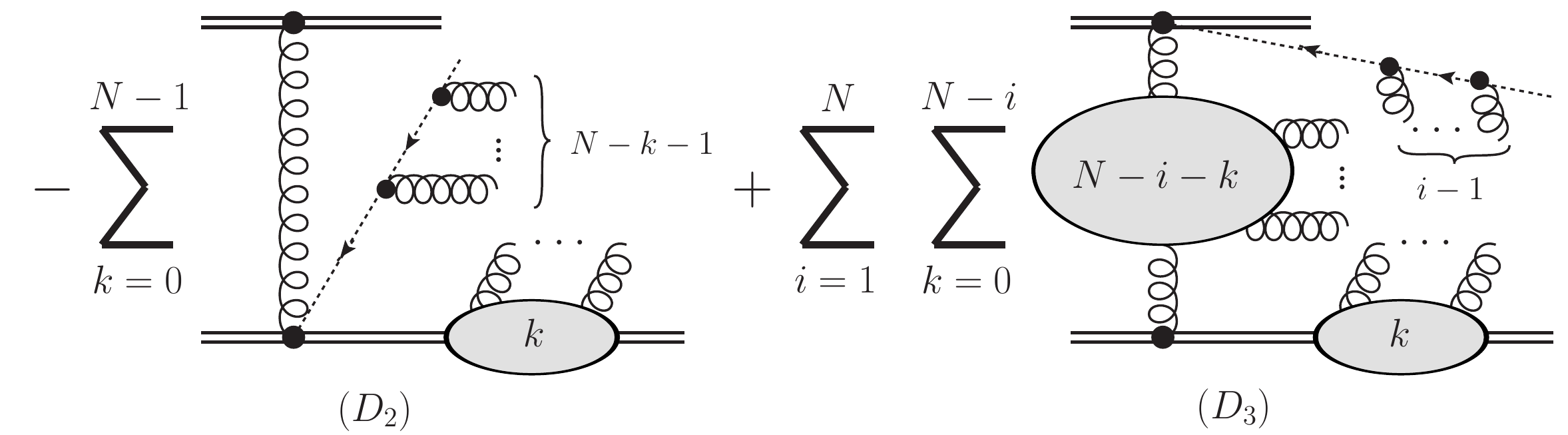} & \centering{}
(\myref)\label{Ward_RR_3}
\tabularnewline
\end{tabular}
\par\end{flushleft}

\noindent Note that the first and the last sum cancel against each
other\begin{equation}
\left(D_{1}\right)+\left(D_{3}\right)=0,\end{equation}
so we are left with the middle sum $\left(D_{2}\right)$. We have
to show that it cancels with the term $\left(C\right)$ of \eqref{Ward_RR_2}.
The last reads

\begin{flushleft}
\begin{tabular}{>{\centering}m{0.87\columnwidth}>{\centering}m{0.05\columnwidth}}
\bigskip{}

\raggedright{}\includegraphics[height=0.08\paperheight]{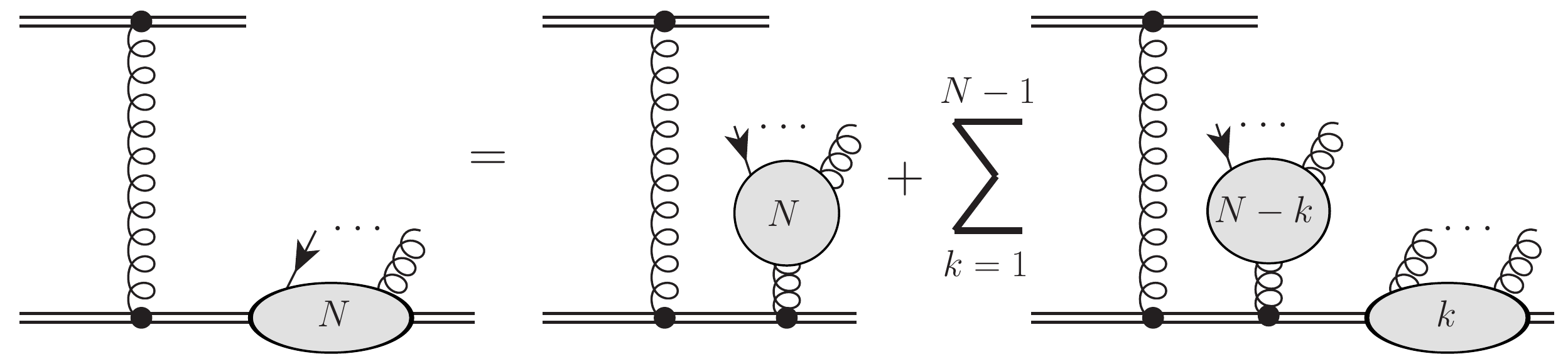} & \centering{}\tabularnewline
\end{tabular}
\par\end{flushleft}

\begin{flushleft}
\begin{tabular}{>{\centering}m{0.87\columnwidth}>{\centering}m{0.05\columnwidth}}
\raggedleft{}\includegraphics[height=0.085\paperheight]{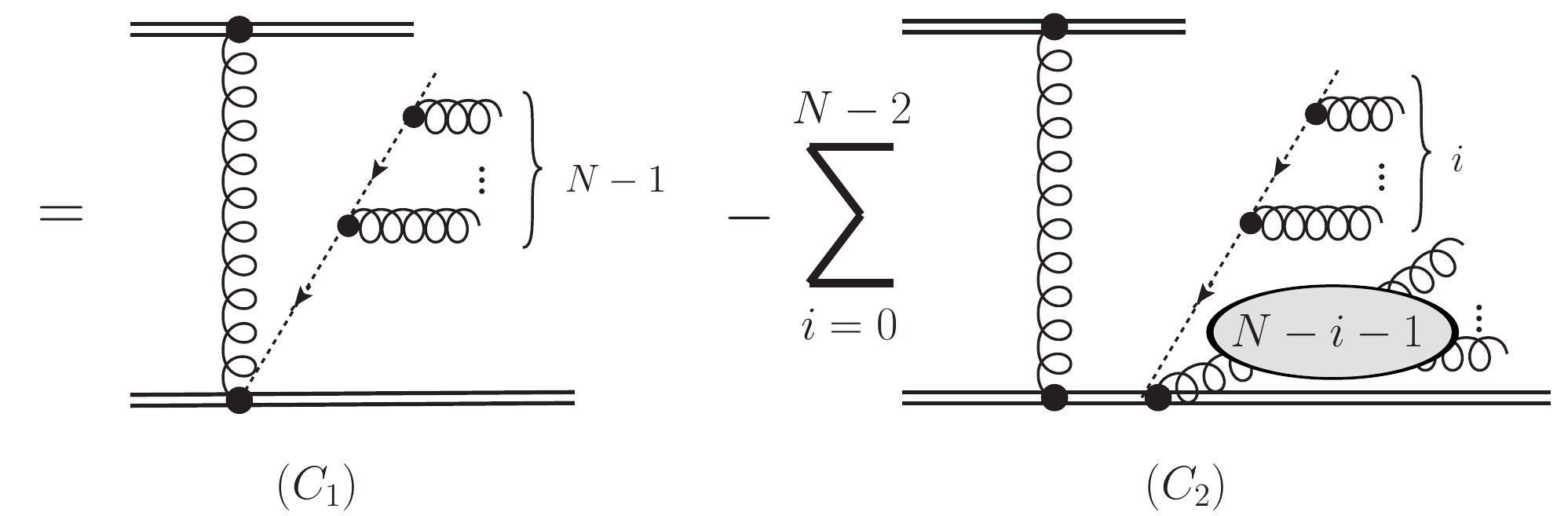} & \centering{}\tabularnewline
\end{tabular}
\par\end{flushleft}

\begin{flushleft}
\begin{tabular}{>{\centering}m{0.87\columnwidth}>{\centering}m{0.05\columnwidth}}
\raggedleft{}\includegraphics[height=0.085\paperheight]{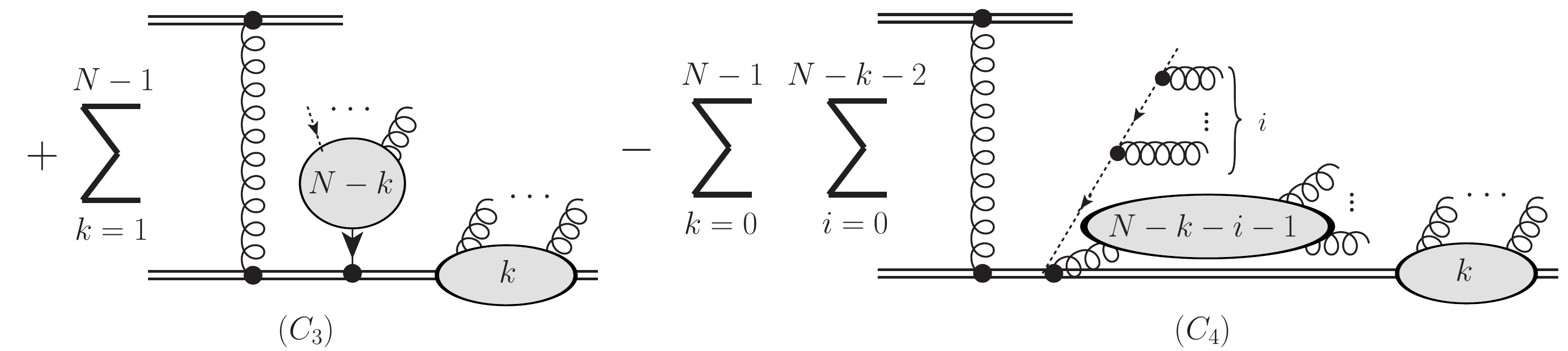} & \centering{} 
(\myref)\label{Ward_RR_5}
\tabularnewline
\end{tabular}
\par\end{flushleft}

\noindent In order to proceed, let us note the following Slavnov-Taylor-like
identity for a gauge link; it follows simply from the momentum conservation,
the form of the gauge-link propagator and the asymmetry of the color-ordered
BRST gluon-ghost vertex

\begin{flushleft}
\begin{tabular}{>{\centering}m{0.87\columnwidth}>{\centering}m{0.05\columnwidth}}
\bigskip{}

\centering{}\includegraphics[height=0.04\paperheight]{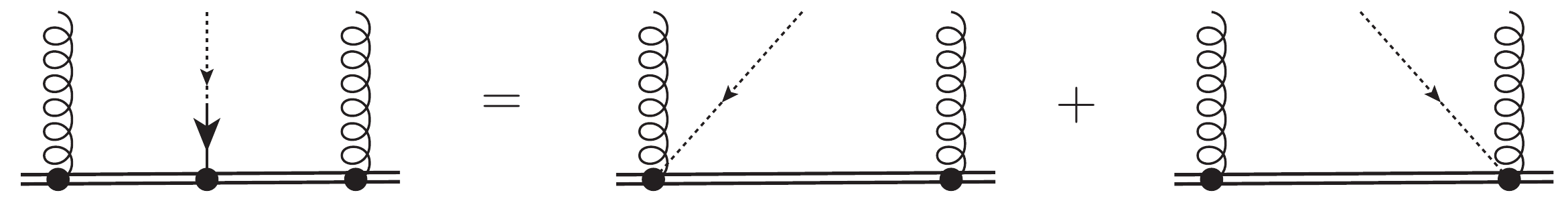} & \centering{}
(\myref)\label{Ward_RR_6}
\tabularnewline
\end{tabular}
\par\end{flushleft}

\noindent Expanding the blob with $k$ legs in the diagram $\left(C_{3}\right)$
according to \eqref{Blob_expansion} and using the above result
we have

\begin{flushleft}
\begin{tabular}{>{\centering}m{0.87\columnwidth}>{\centering}m{0.05\columnwidth}}
\bigskip{}

\centering{}\includegraphics[height=0.09\paperheight]{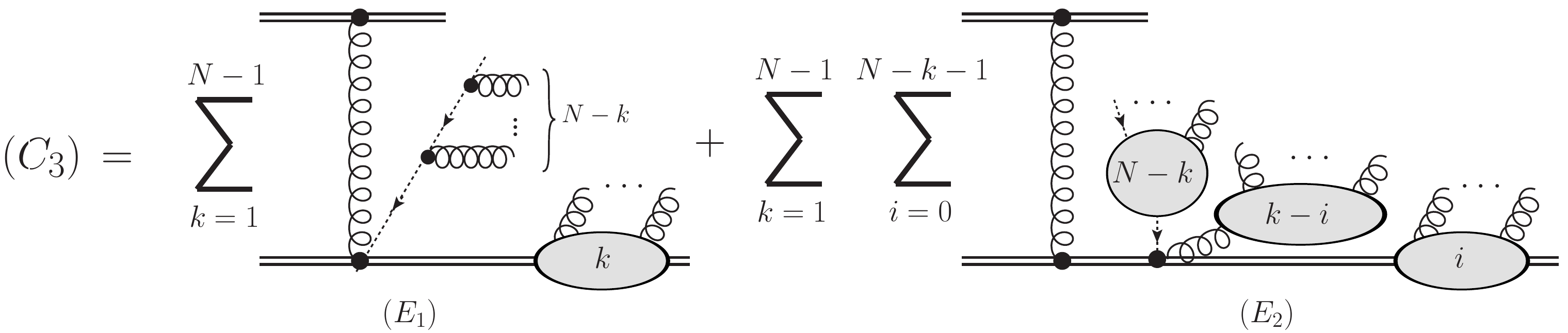} & \centering{}
(\myref)\label{Ward_RR_7}
\tabularnewline
\end{tabular}
\par\end{flushleft}

\noindent After reshuffling the sums, the second term $\left(E_{2}\right)$
cancels with $\left(C_{4}\right)$ due to \eqref{ST_axial}. Finally,
it is easy to see that \begin{equation}
\left(E_{1}\right)+\left(C_{1}\right)+\left(D_{2}\right)=0.\end{equation}

We have shown that all the terms on the r.h.s of \eqref{Ward_RR_2}
cancel against each other and thus the Ward identity is indeed fulfilled.
Similar proofs for more gauge links are also possible, but very cumbersome
to carry out explicitly. For three and four gauge links we have checked
the Ward identities using explicit results presented in section \ref{sec:Examples}.

Let us note that the proofs are in a sense purely algebraic and they
do not refer to the form of the off-shell momenta or gauge link directions,
the only property that is relevant here is their mutual transversality.

\section{Examples}

\label{sec:Examples}

\subsection{Gauge invariant amplitude for $g^{*}g^{*}g$}

\label{sub:Examp_RRG}

Let us start with a simple example of a gauge invariant matrix element
for the following process\begin{equation}
g^{*}\left(k_{A}\right)g^{*}\left(k_{B}\right)\rightarrow g\left(p\right).\label{eq:RRG_1}\end{equation}
The momentum conservation is $k_{A}+k_{B}=p$ with $p^{2}=0$. We
choose the Feynman gauge and polarization vectors to be $\varepsilon_{A}$,
$\varepsilon_{B}$, $\varepsilon$ satisfying $\varepsilon_{A}\cdot k_{A}=0$,
$\varepsilon_{B}\cdot k_{B}=0$, $\varepsilon\cdot p=0$. Let the
colors of the gluons be $c_{A}$, $c_{B}$ and $c$ respectively.

\begin{figure}
\begin{centering}
\includegraphics[width=0.9\textwidth]{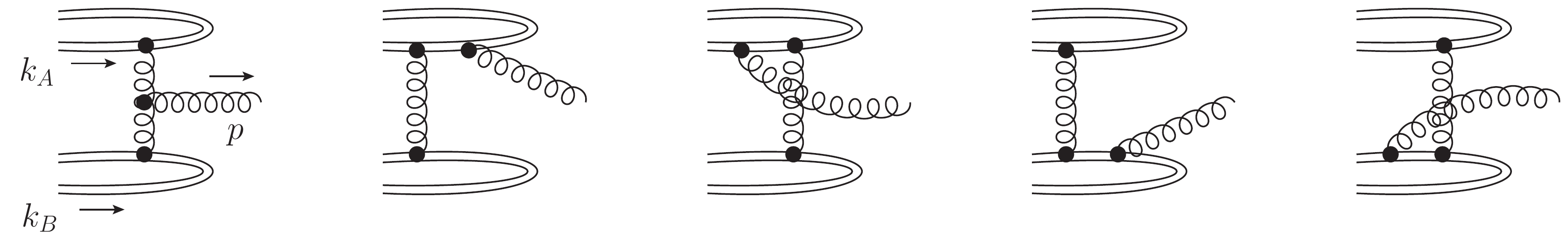}
\par\end{centering}

\caption{\label{fig:RRG_diags} \small The diagrams for the gauge invariant process
$g^{*}\left(k_{A}\right)g^{*}\left(k_{B}\right)\rightarrow g\left(p\right)$.
The momentum flow is displayed for the first diagram only.}
\end{figure}

The relevant Feynman diagrams are displayed in figure~\ref{fig:RRG_diags}.
Using our Feynman rules we get \begin{multline}
\tilde{\mathcal{M}}_{\varepsilon_{A}\varepsilon_{B}}\left(\varepsilon\right)=\varepsilon_{A}^{\mu}\varepsilon_{B}^{\nu}\varepsilon^{\gamma}\frac{1}{k_{A}^{2}k_{B}^{2}}\, f^{c_{A}c_{B}c}V_{\mu\beta\gamma}\left(k_{A},k_{B},-p\right)\\
-2ig\,\frac{\varepsilon_{A}\cdot\varepsilon_{B}\,\varepsilon\cdot\varepsilon_{A}}{k_{B}^{2}\, p\cdot\varepsilon_{A}}\left[\mathrm{Tr}\left(t^{c_{A}}t^{c_{B}}t^{c}\right)-\mathrm{Tr}\left(t^{c_{A}}t^{c}t^{c_{B}}\right)\right]\\
-2ig\,\frac{\varepsilon_{A}\cdot\varepsilon_{B}\,\varepsilon\cdot\varepsilon_{B}}{k_{A}^{2}\, p\cdot\varepsilon_{B}}\left[\mathrm{Tr}\left(t^{c_{B}}t^{c_{A}}t^{c}\right)-\mathrm{Tr}\left(t^{c_{B}}t^{c}t^{c_{A}}\right)\right]\\
=igf^{c_{A}c_{B}c}\bigg\{\frac{-i\varepsilon_{A}^{\mu}\varepsilon_{B}^{\nu}\varepsilon^{\gamma}V_{\mu\beta\gamma}\left(k_{A},k_{B},-p\right)}{k_{A}^{2}k_{B}^{2}}+\\
\frac{\varepsilon_{A}\cdot\varepsilon_{B}\,\varepsilon\cdot\varepsilon_{A}}{k_{B}^{2}\, p\cdot\varepsilon_{A}}-\frac{\varepsilon_{A}\cdot\varepsilon_{B}\,\varepsilon\cdot\varepsilon_{B}}{k_{A}^{2}\, p\cdot\varepsilon_{B}}\bigg\}\label{eq:RRG}\end{multline}
where $V_{\alpha\beta\gamma}^{abc}\left(k_{1},k_{2},k_{3}\right)=f^{abc}V_{\alpha\beta\gamma}\left(k_{1},k_{2},k_{3}\right)$
with\begin{equation}
V_{\alpha\beta\gamma}\left(k_{1},k_{2},k_{3}\right)=-g\left[\eta_{\alpha\beta}\left(k_{1}-k_{2}\right)_{\gamma}+\eta_{\beta\gamma}\left(k_{2}-k_{3}\right)_{\alpha}+\eta_{\gamma\alpha}\left(k_{3}-k_{1}\right)_{\beta}\right]\label{eq:V3b}\end{equation}
 is the three-gluon coupling. It is easy to see that the above result
recovers the Lipatov's RRP vertex \cite{Antonov:2004hh}, provided
$\varepsilon_{A}$ and $\varepsilon_{B}$ are replaced by $n_{-}$
and $n_{+}$ where $n_{\pm}=\left(1,0,0,\mp1\right)$ define the {}``plus''
and {}``minus'' light-cone directions. Actually, to get the precise
equality, our eq.~(\ref{eq:RRG}) has to be multiplied by the inverse
propagators of the off-shell gluons, i.e. $k_{A}^{2}k_{B}^{2}$. We
will see in the next subsection, that such one-to-one correspondence
between our matrix elements and the Lipatov vertices does not hold
for more off-shell gluons.

There are two color ordered amplitudes $\tilde{\mathcal{M}}_{\varepsilon_{A}\varepsilon_{B}}^{\left(c_{A}c_{B}c\right)}$
and $\tilde{\mathcal{M}}_{\varepsilon_{A}\varepsilon_{B}}^{\left(c_{B}c_{A}c\right)}$
that contribute to eq.~(\ref{eq:RRG}). They are easy to read out
--- they are the coefficients of the color traces. We have checked
that they match the result obtained automatically form the $\mathtt{OGIME}$
program discussed in section~\ref{sec:OGIME}.

\subsection{Gauge invariant amplitude for $g^{*}g^{*}g^{*}g$}

\label{sec:RRRG}

We turn now to the gauge-invariant amplitude for the process with
three off-shell gluons and one on-shell. Let us assign the momenta
as follows\begin{equation}
g^{*}\left(k_{A}\right)g^{*}\left(k_{C}\right)\rightarrow g^{*}\left(k_{B}\right)g\left(p\right),\end{equation}
and assign the color quantum numbers to be $c_{A}$, $c_{C}$, $c_{B}$,
$c$ respectively. The momentum conservation reads $k_{A}+k_{C}=k_{B}+p$
with $p^{2}=0$. Let us assume, that the on-shell gluon has the polarization
vector $\varepsilon$ whereas the off-shell gluons are assigned the
polarization vectors $\varepsilon_{A}$, $\varepsilon_{B}$, $\varepsilon_{C}$
transverse to the momenta $k_{A}$, $k_{B}$, $k_{C}$ accordingly.

\begin{figure}
\begin{centering}
\includegraphics[width=0.9\textwidth]{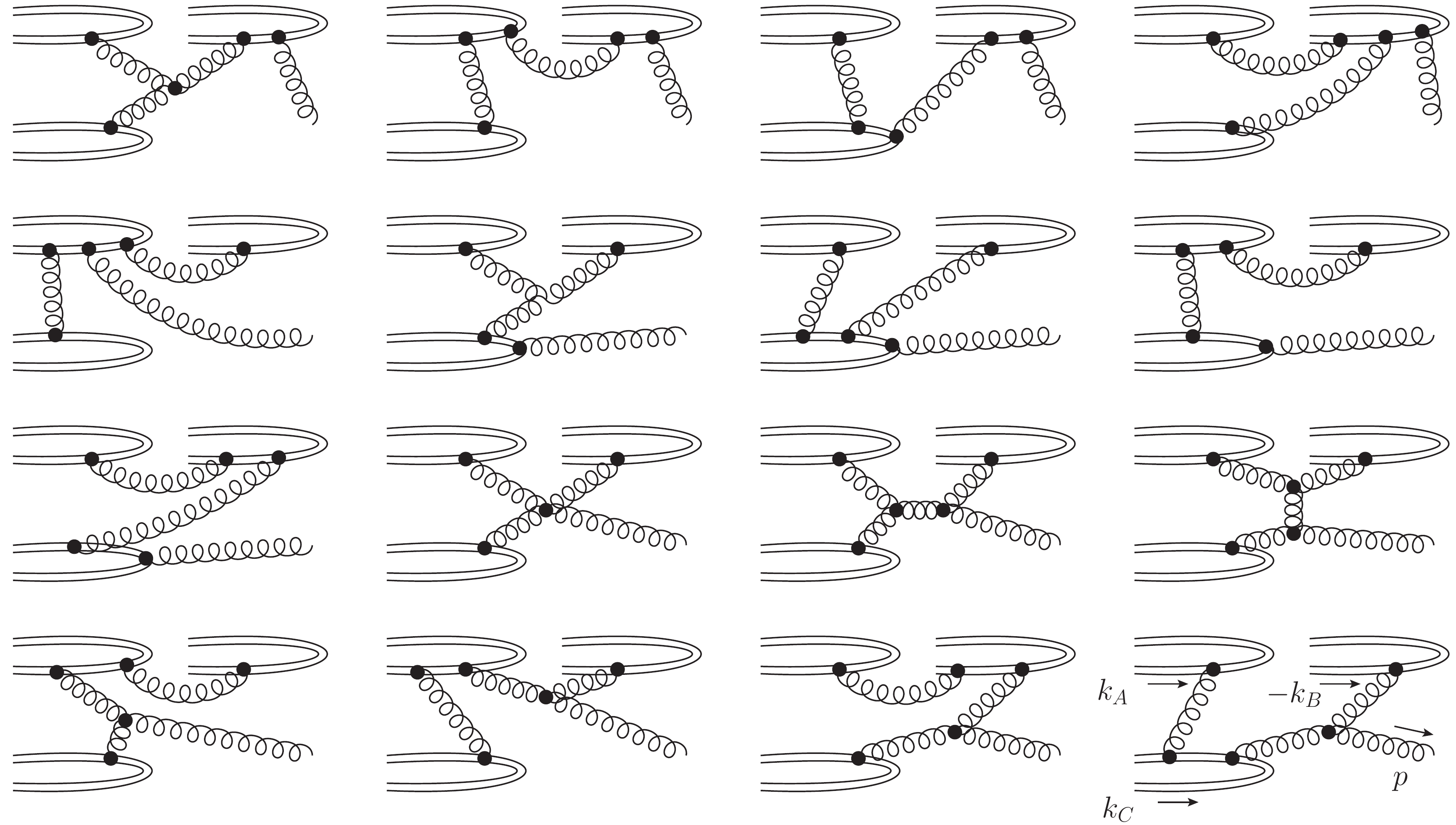}
\par\end{centering}

\caption{\label{fig:RRRG_diags} \small The color-ordered diagrams for the process
$g^{*}\left(k_{A}\right)g^{*}\left(k_{C}\right)\rightarrow g^{*}\left(k_{B}\right)g\left(p\right)$
and color order $\left(c_{A},c_{B},c,c_{C}\right)$. The momentum
flow is displayed in the last diagram only.}
\end{figure}

Let us work in the Feynman gauge and consider the color-ordered amplitude
with the color order $\left(c_{A},c_{B},c,c_{C}\right)$. There are
16 color ordered diagrams shown in figure~\ref{fig:RRRG_diags}. It
is straightforward to calculate them using the color ordered Feynman
rules given in section \ref{sec:Feynman_rules} (and the ones for
standard QCD listed e.g.\ in \cite{vanHameren:2012uj}) . The relevant
gauge links are defined using $\varepsilon_{A}$, $\varepsilon_{B}$,
$\varepsilon_{C}$ vectors.

The result from the Feynman diagrams has been cross-checked with the
$\mathtt{OGIME}$ program. We append the analytic result to the paper
in a text file \href{file:./RRRG.txt}{RRRG.txt}. The result is given
in the most general form, when $\varepsilon_{A}\neq\varepsilon_{B}\neq\varepsilon_{C}$.
If any of those vectors coincide they should be light-like (see section
\ref{sec:formal_develop}). We have checked explicitly that our result
satisfies the Ward identity\begin{equation}
\mathcal{M}_{\varepsilon_{A}\varepsilon_{B}\varepsilon_{C}}^{\left(c_{A},c_{B},c,c_{C}\right)}\left(p\right)=0.\end{equation}
The other color ordered amplitudes can be obtained by suitable exchanges
of the pairs $\left(k_{X},\varepsilon_{X}\right)$, $X=A,B,C$. The
full result for the squared amplitude is just given by the sum of
color ordered amplitudes squared (let us recall that it is not a general
rule and holds for at most five external legs, see e.g.\ \cite{Mangano:1990by}).

Let us stress that the above result corresponds to the known result
for RRRG vertex given in \cite{Braun:2006sk} only for a special
case, such that $\varepsilon_{A}=n_{+}$, $\varepsilon_{B}=\varepsilon_{C}=n_{-}$
and $k_{A}=x_{A}n_{-}+k_{A\, T}$, $k_{B}=x_{B}n_{+}+k_{B\, T}$,
$k_{C}=x_{C}n_{+}+k_{C\, T}$ with $n_{\pm}$ defined in the previous
section. With the above choice of the kinematics the diagrams with
a gluon joining the Wilson lines with momenta $k_{B}$ and $k_{C}$
vanish. The precise relation to the result of \cite{Braun:2006sk}
is the following: our result should be multiplied by $ik_{A}^{2}k_{B}^{2}k_{C}^{2}/6$.
The origin of inverse propagators is obvious, while the factor of
6 comes entirely from our convention of defining the $\mathfrak{R}$
operator (there are three of them, each contributes a factor of 2).
Namely, our definition is such that when the on-shell limit is taken
(after multiplying by inverse off-shell propagators), we obtain the
standard text-book result. This is also the case for the Lipatov vertices
listed in \cite{Antonov:2004hh}, but not for \cite{Braun:2006sk}.
This is evident if we compare the RRP vertices presented in both papers
--- they differ by a factor of 4.

We underline, that our result is more general than the one listed
in \cite{Braun:2006sk}, as it allows for arbitrary {}``orientation''
of reggeized gluons. The special cases can be obtained by imposing
additional restrictions on the $\varepsilon_{A}$, $\varepsilon_{B}$,
$\varepsilon_{C}$ vectors.

\subsection{Gauge invariant amplitude for $g^{*}g^{*}g^{*}g^{*}g$}

\label{sub:RRRRG}

Finally, let us turn to the process\begin{equation}
g^{*}\left(k_{A}\right)g^{*}\left(k_{D}\right)\rightarrow g^{*}\left(k_{B}\right)g^{*}\left(k_{C}\right)g\left(p\right).\end{equation}
The polarization vectors are respectively $\varepsilon_{A}$, $\varepsilon_{D}$,
$\varepsilon_{B}$, $\varepsilon_{C}$, $\varepsilon$ and the colors
$c_{A}$, $c_{D}$, $c_{B}$, $c_{C}$, $c$.

The calculation using the Feynman diagrams is rather lengthy even
for color-ordered amplitude, therefore we have calculated this amplitude
using $\mathtt{OGIME}$ program only. The result is appended to the
paper in a text file \href{file:./RRRRG.txt}{RRRRG.txt} for the most
general form, when $\varepsilon_{A}\neq\varepsilon_{B}\neq\varepsilon_{C}\neq\varepsilon_{D}$.
If any of those vectors coincide they should be light-like (see section
\ref{sec:formal_develop}). We have explicitly checked that the result
satisfies the Ward identity.

We note, that for certain choice of polarization vectors $\varepsilon_{A}=n_{+}$,
$\varepsilon_{B}=n_{-}$, $\varepsilon_{C}=n_{-}$, $\varepsilon_{D}=n_{-}$
(where $n_{\pm}$ were defined in the previous section) the result
corresponds to $RRRRP$ Lipatov's vertex, calculated in \cite{Braun:2012gs}.
Similar as previously, the result we present is however more general
as it allows for any {}``orientation'' for reggeized gluons.

\section{Gauge invariant decompositions}

\label{sec:GIdecomp}

Let us now discuss another interesting application of gauge invariant
off-shell matrix elements. As we have indicated it already a few times,
they can be built using almost arbitrary gauge link directions. Below,
we will use this fact to decompose an ordinary gauge invariant amplitude
into gauge invariant off-shell sub-amplitudes.

Consider first a standard tree-level off-shell current $J^{\mu}\left(\varepsilon_{1},\ldots,\varepsilon_{N};k\right)$
in the Feynman gauge, where $\varepsilon_{1},\ldots,\varepsilon_{N}$
are polarization vectors of $N$ on-shell gluons with momenta $k_{1},\ldots,k_{N}$
respectively and $k=k_{1}+\ldots+k_{N}$ (we omit color indices for
brevity). Let us note that we may write\begin{equation}
J^{\mu}\left(\varepsilon_{1},\ldots,\varepsilon_{N};k\right)=\sum_{i=0}^{2}J_{\nu}\left(\varepsilon_{1},\ldots,\varepsilon_{N};k\right)\epsilon_{i}^{\nu}\left(k\right)\epsilon_{i}^{\mu}\left(k\right)d_{i}\left(k\right).\label{eq:GIdec_1}\end{equation}
The auxiliary (in general complex) four-vectors $\epsilon_{i}\left(k\right)$
are defined in such a way that\begin{equation}
k\cdot\epsilon_{i}\left(k\right)=0,\,\,\,\,\,\,\,\epsilon_{i}\left(k\right)\cdot\epsilon_{j}\left(k\right)=d_{i}\left(k\right)\delta_{ij},\label{eq:GIdecomp_2a}\end{equation}
\begin{equation}
\sum_{i=0}^{2}\epsilon_{i}^{\nu}\left(k\right)\epsilon_{i}^{\mu}\left(k\right)d_{i}\left(k\right)=g^{\mu\nu}-\frac{k^{\mu}k^{\nu}}{k^{2}},\label{eq:GIdec_2b}\end{equation}
where\begin{equation}
d_{0}\left(k\right)=\begin{cases}
1 & k^{2}<0\\
-1 & k^{2}>0\end{cases},\,\,\,\,\,\,\, d_{1}\left(k\right)=d_{2}\left(k\right)=-1.\label{eq:GIdec_3}\end{equation}
The identity (\ref{eq:GIdec_1}) follows because of the current conservation
$J^{\mu}\left(\varepsilon_{1},\ldots,\varepsilon_{N};k\right)k_{\mu}=0$.
The auxiliary four vectors $\epsilon_{i}\left(k\right)$ can be easily
constructed for any off-shell four momentum $k$. For instance, if
$k$ is space-like we go to a frame where $k=\left(0,0,0,\sqrt{\left|k^{2}\right|}\right)$
and define $\epsilon_{0}\left(k\right)=\left(1,0,0,0\right)$, $\epsilon_{1}\left(k\right)=\left(0,\sin\phi,\cos\phi,0\right)$,
$\epsilon_{2}\left(k\right)=\left(0,-\cos\phi,\sin\phi,0\right)$
and eventually go back to the original frame. Obviously, the four-vectors
$\epsilon_{i}\left(k\right)$ are not unique.

The off-shell current $J^{\mu}\left(\varepsilon_{1},\ldots,\varepsilon_{N};k\right)$
is not gauge invariant. Thanks to the fact that $\epsilon_{i}\left(k\right)\cdot k=0$
we may however define a \textit{gauge invariant off-shell current}
as follows\begin{multline}
\tilde{J}_{i}^{\mu}\left(\varepsilon_{1},\ldots,\varepsilon_{N};k\right)=\tilde{J}_{i}\left(\varepsilon_{1},\ldots,\varepsilon_{N};k\right)\epsilon_{i}^{\mu}\left(k\right)d_{i}\left(k\right)\\
=\left[-\frac{1}{k^{2}}\, J_{\nu}\left(\varepsilon_{1},\ldots,\varepsilon_{N};k\right)\epsilon_{i}^{\nu}\left(k\right)+G_{i}\left(\varepsilon_{1},\ldots,\varepsilon_{N};k\right)\right]\epsilon_{i}^{\mu}\left(k\right)d_{i}\left(k\right),\label{eq:GIdec_4-1}\end{multline}
where \begin{equation}
\tilde{J}_{i}\left(\varepsilon_{1},\ldots,\varepsilon_{N};k\right)\overset{*}{=}\left\langle k_{1},\varepsilon_{1};\ldots;k_{N},\varepsilon_{N}\right|\mathcal{R}_{\epsilon_{i}}\left(k\right)\left|0\right\rangle \label{eq:GIdec_4a}\end{equation}
and $G_{i}\left(\varepsilon_{1},\ldots,\varepsilon_{N};k\right)$
are contributions coming from additional emissions from the Wilson
line residing in $\mathcal{R}_{\epsilon_{i}}\left(k\right)$ (they
are generalization of the {}``gauge-restoring amplitude'' $\mathcal{W}$
from section \ref{sec:gauge_restoring_amp}. The above equation leads
to the following decomposition of the standard off-shell current into
gauge invariant and gauge non-invariant pieces\begin{equation}
J^{\mu}\left(\varepsilon_{1},\ldots,\varepsilon_{N};k\right)=-\sum_{i=0}^{2}\left[k^{2}\,\tilde{J}_{i}^{\mu}\left(\varepsilon_{1},\ldots,\varepsilon_{N};k\right)-G_{i}\left(\varepsilon_{1},\ldots,\varepsilon_{N};k\right)\epsilon_{i}^{\mu}\left(k\right)d_{i}\left(k\right)\right].\label{eq:GIdec_5}\end{equation}

Consider now a particular example, namely a color-ordered on-shell
amplitude for the four gluon process $gggg\rightarrow0$. It can be
expressed in terms of color-ordered currents as follows\begin{multline}
\mathcal{M}^{\left(1,2,3,4\right)}\left(\varepsilon_{1},\varepsilon_{2},\varepsilon_{3},\varepsilon_{4}\right)=J_{\mu}^{\left(1,2\right)}\left(\varepsilon_{1},\varepsilon_{2};k_{12}\right)\frac{i}{k_{12}^{2}}\, J^{\mu\,\left(3,4\right)}\left(\varepsilon_{3},\varepsilon_{4};k_{34}\right)\\
+J_{\mu}^{\left(4,1\right)}\left(\varepsilon_{1},\varepsilon_{4};k_{14}\right)\frac{i}{k_{14}^{2}}\, J^{\mu\,\left(2,3\right)}\left(\varepsilon_{2},\varepsilon_{3};k_{23}\right)+V_{4}^{\left(1,2,3,4\right)}\left(\varepsilon_{1},\varepsilon_{2},\varepsilon_{3},\varepsilon_{4}\right),\label{eq:GIdec_6}\end{multline}
where \begin{equation}
V_{4}^{\left(1,2,3,4\right)}\left(\varepsilon_{1},\varepsilon_{2},\varepsilon_{3},\varepsilon_{4}\right)=\frac{g^{2}}{2}\, V_{4}^{\alpha_{1}\alpha_{2}\alpha_{3}\alpha_{4}}\varepsilon_{1\,\alpha_{1}}\varepsilon_{2\,\alpha_{2}}\varepsilon_{3\,\alpha_{3}}\varepsilon_{4\,\alpha_{4}}\label{eq:GIdec_6b}\end{equation}
is the color-ordered four gluon coupling. The color-ordered objects,
with the color order indicated in the superscript parentheses were
defined in section \ref{sec:gauge_restoring_amp}. In what follows,
we drop the arguments of color-ordered objects; such a shortcut notation
does not lead to any confusion here. Using the gauge invariant currents
(\ref{eq:GIdec_4-1}) we may write\begin{equation}
\mathcal{M}^{\left(1,2,3,4\right)}=i\left(k_{12}^{2}\,\tilde{J}^{\left(1,2\right)}\cdot\tilde{J}^{\left(3,4\right)}+k_{14}^{2}\,\tilde{J}^{\left(4,1\right)}\cdot\tilde{J}^{\left(2,3\right)}+\tilde{V}_{4}^{\left(1,2,3,4\right)}\right),\label{eq:GIdec_7}\end{equation}
where the gauge invariant (color-ordered) four-gluon vertex is defined
as\begin{multline}
\tilde{V}_{4}^{\left(1,2,3,4\right)}=-iV_{4}^{\left(1,2,3,4\right)}-G^{\left(1,2\right)}\cdot\tilde{J}^{\left(3,4\right)}-\tilde{J}^{\left(1,2\right)}\cdot G^{\left(3,4\right)}+\frac{1}{k_{12}}\, G^{\left(1,2\right)}\cdot G^{\left(3,4\right)}\\
-G^{\left(4,1\right)}\cdot\tilde{J}^{\left(2,3\right)}-\tilde{J}^{\left(4,1\right)}\cdot G^{\left(2,3\right)}+\frac{1}{k_{14}}\, G^{\left(4,1\right)}\cdot G^{\left(2,3\right)}.\label{eq:GIdec_8}\end{multline}
The scalar product in the expressions above is defined as \begin{equation}
A\cdot B=\sum_{i=0}^{2}A_{i}B_{i}d_{i}.\end{equation}

Let us summarize what we have done. We have used the matrix elements
of Wilson lines to decompose a gauge invariant on-shell amplitude
(here we have used a simple example of four-leg amplitude) to manifestly
gauge invariant objects. Such decomposition is however not unique,
as the gauge invariant objects depend on the choice of four-vectors
$\epsilon_{i}$. Similar decompositions are possible also for more
complicated objects. 

The practical applications of such decompositions are under study.
At this point, as an example, let us turn the attention to the ref.~\cite{Richter-Was:184418},
where a decomposition of an on-shell amplitude to $ggg^{*}$ subprocess
(and the reminder) was used in order to study the spin effects in
the QCD evolution. There, however, the gauge-violating part was abandoned
as it was a non-leading-logarithmic contribution.

\section{Discussion and outlook}

\label{sec:summary}

We start by summarizing the main points of the paper. The notion {}``off-shell
amplitude'' is motivated by the high-energy factorization and refers
to an ordinary scattering amplitude, where some of the gluonic legs
\textit{are not} taken on-shell, \textit{plus} contributions needed
to maintain the gauge invariance. We have defined such off-shell amplitudes
in eq.~(\ref{eq:ME_deltas-1-1-1}) via matrix elements of the Fourier
transforms of straight infinite Wilson line operators. The {}``polarization
vectors'' of off-shell gluons correspond (in the Feynman gauge) to
the directions of the Wilson lines. We have implemented the method
in the FORM program, called $\mathtt{OGIME}$, and tested it for certain
processes involving several off-shell and on-shell gluons. In the
present work only tree-level gluonic matrix elements were studied.

As there are many applications of Wilson lines in the high-energy
literature, let us briefly compare their various instances to the
present one. Let us start with the Lipatov's effective action \cite{Lipatov:1995pn,Antonov:2004hh}.
The Lipatov's vertices are formulated in terms of two auxiliary fields
$A_{+}$ and $A_{-}$. Due to the equations of motion they are related
to the Wilson lines defined in terms of two light-like four vectors
$n_{+}$, $n_{-}$ defining the {}``plus'' and {}``minus'' light
cone components. The Wilson lines are however not infinite there,
but they start at certain fixed space-time position and span to infinity.
If we defined the matrix elements in terms of the $\mathcal{R}$ operators
with the Wilson lines defined using only $n_{\pm}$ four vectors we
would recover the Lipatov's vertices. It was also demonstrated explicitly
in section \ref{sec:Examples}. In \cite{Balitsky:1995ub,Balitsky:2001gj}
and \cite{Caron-Huot:2013fea} the Wilson line operators are infinite,
but are explicitly taken to have {}``plus'' or {}``minus'' light-cone
component set to zero. This corresponds to our integrated $\mathfrak{R}$
operators, c.f. eq.~(\ref{eq:Rphys_def}), with gauge link direction
again defined to be along the {}``plus'' or {}``minus'' light-cone
vector. The present approach is more general --- the gauge link directions
need not to be constrained to the {}``plus'' or {}``minus'' light-cone
vectors and there are additional contributions related to one gluon
exchanges between the Wilson lines. 

Let us also compare the present approach to the one of ref.~\cite{vanHameren:2012if}.
There, also a set of eikonal Feynman rules was introduced, using however
completely different method. Again, for our $\mathfrak{R}$ operators
taken to be along the hadrons momenta the results from both methods
coincide. However, at the moment the method \cite{vanHameren:2012if}
takes into account at most two off-shell gluons within high energy
kinematics described in section \ref{sec:gauge_restoring_amp}.

One of the main purposes of this work was to provide a practical formulation
for off-shell amplitudes. We may claim that we have reach that goal,
as the relevant program, the Feynman rules, as well as explicit calculations
were presented. Since the program calculates the amplitudes analytically,
it obviously has certain limitations; already for five external legs
with arbitrary polarization vectors the results are rather lengthy.
Therefore, as far as the future applications for analytic calculations
are concerned it is rather more interesting to include other Standard
Model fields in the program. This is planned for the near future.

\acknowledgments

The author is grateful for many useful discussions and reading the
manuscript to A.~van~Hameren and K.~Kutak. Useful comments and
questions were also given by L.~Motyka, M.~Sadzikowski and W.~Schafer,
to whom the author is grateful as well.

The work was partially supported by the Polish National Science Center
grants \\ LIDER/02/35/\mbox{L-2}/10/NCBiR/2011 and \mbox{DEC-2011}/01/B/ST2/03643.

The Feynman diagrams were drawn using JaxoDraw program \cite{Binosi200476}.

\appendix

\section{Direct calculation of gauge invariant $g^{*}gg$ matrix element}

\label{sec:App_RGG}

In order to illustrate how the $\mathtt{OGIME}$ program calculates
the matrix elements let us present a sample calculation. It uses only
very basics of the Quantum Field Theory.

The precise definition of the process is as follows \begin{equation}
g^{*}\left(k_{A}\right)g\left(k_{1}\right)\rightarrow g\left(k_{2}\right),\end{equation}
where the gluons have colors $c_{A}$, $a_{1}$, $a_{2}$ and polarizations
$\varepsilon_{A}$, $\varepsilon_{1}^{\lambda_{1}}$, $\varepsilon_{2}^{\lambda_{2}}$
respectively. Here, unlike in the main text we explicitly indicate
the polarization projections $\lambda_{1}$, $\lambda_{2}$.

For the purpose of the explicit derivations let us define the necessary
field operators in the Heisenberg picture. We have for a gluon field\begin{equation}
A_{\mu}^{c}\left(x\right)=\int\widetilde{dq}\,\frac{1}{\sqrt{2E_{q}}}\,\sum_{\lambda}\left[\hat{a}_{c}^{\lambda}\left(q\right)\varepsilon_{\mu}^{\lambda}\left(q\right)e^{-iq\cdot x}+\hat{a}_{c}^{\lambda\dagger}\left(q\right)\varepsilon_{\mu}^{\lambda*}\left(q\right)e^{iq\cdot x}\right],\end{equation}
where $E_{q}=\sqrt{q^{2}+\vec{q}^{2}}$ and the hatted quantities
are creation/annihilation operators with the commutation relations\begin{equation}
\left[\hat{a}_{c}^{\lambda}\left(q\right),\hat{a}_{c'}^{\lambda'\dagger}\left(q'\right)\right]=\left(2\pi\right)^{3}\delta^{3}\left(\vec{q}-\vec{q}'\right)\delta_{\lambda\lambda'}\delta_{cc'}.\end{equation}
We introduce the following shorthand notation for the integration
measures\begin{equation}
\frac{d^{4}p}{\left(2\pi\right)^{4}}\equiv\widehat{dp},\,\,\,\,\,\,\,\,\,\,\,\frac{d^{3}p}{\left(2\pi\right)^{3}}\equiv\widetilde{dp}.\end{equation}

A one-particle on-shell state with momentum $q$, color $c$ and polarization
vector $\varepsilon^{\lambda}\left(q\right)$ is defined as\begin{equation}
\left|q,\lambda,c\right\rangle \equiv\left|q,\varepsilon^{\lambda}\left(q\right),c\right\rangle =\sqrt{2E_{q}}\,\hat{a}_{c}^{\lambda\dagger}\left(q\right)\left|0\right\rangle .\end{equation}
The relevant interaction part of the Yang-Mills action is (skipping
ghost and gauge-fixing parts)

\begin{equation}
S_{I}=\int d^{4}x\,\left(gV_{ggg}\left(x\right)+g^{2}V_{gggg}\left(x\right)\right),\label{eq:App_SI}\end{equation}
where\begin{equation}
V_{ggg}\left(x\right)=f_{abc}\partial_{\mu}A_{\nu}^{a}\left(x\right)\, A^{b\,\mu}\left(x\right)\, A^{c\,\nu}\left(x\right),\end{equation}
\begin{equation}
V_{gggg}\left(x\right)=-\frac{1}{4}\,\left(f_{abc}A_{\mu}^{b}\left(x\right)A_{\nu}^{c}\left(x\right)\right)^{2},\end{equation}
For further purposes, let us write $V_{ggg}$ interactions in the
form\begin{equation}
V_{ggg}\left(x\right)=f_{abc}\int\widehat{dp}\int d^{4}y\,\,\left(-i\right)p_{\mu}e^{-i\left(x-y\right)\cdot p}\, A_{\nu}^{a}\left(y\right)A^{b\,\mu}\left(x\right)\, A^{c\,\nu}\left(x\right).\end{equation}

The matrix element we are to calculate is \begin{equation}
\mathfrak{M}\left(\varepsilon_{A};\varepsilon_{1},\varepsilon_{2}\right)\overset{*}{=}\int d^{4}x\, e^{ix\cdot k_{A}}\left\langle k_{1},\lambda_{1},a_{1}\right|\mathrm{Tr}\left\{ \frac{1}{\pi g}t^{c_{A}}\left[x\right]_{\varepsilon_{A}}\right\} \left|k_{2},\lambda_{2},a_{2}\right\rangle .\end{equation}
First, we use the Gell-Mann-Low formula what accounts in inserting
the exponential of the Yang-Mills interaction (\ref{eq:App_SI}),
$\exp\left(-iS_{I}\right)$, switching to the interaction picture
(in what follows we shall not indicate this explicitly) and time ordering
the fields. Taking the connected diagrams only we are left with\begin{equation}
\mathfrak{M}\left(\varepsilon_{A};\varepsilon_{1},\varepsilon_{2}\right)=\int d^{4}x\, e^{ix\cdot k_{A}}\left(M_{1}+M_{2}\right)=\tilde{M}_{1}+\tilde{M}_{2},\end{equation}
where the tildes denote the Fourier transforms with respect to $x$
and\begin{equation}
M_{1}=-\frac{i^{2}g}{\pi}\,\mathrm{Tr}\left(t^{c_{A}}t^{b}\right)\int_{-\infty}^{\infty}ds\int d^{4}y\,\left\langle k_{1},\lambda_{1},a_{1}\right|\,\mathcal{T}\,\varepsilon_{A}\cdot A^{b}\left(x+s\varepsilon_{A}\right):V_{ggg}\left(y\right):\left|k_{2},\lambda_{2},a_{2}\right\rangle ,\end{equation}
\begin{multline}
M_{2}=\frac{i^{2}g}{\pi}\,\mathrm{Tr}\left(t^{c_{A}}t^{b}t^{b'}\right)\int_{-\infty}^{\infty}ds\int_{-\infty}^{s}ds'\,\\
\left\langle k_{1},\lambda_{1},a_{1}\right|\,\mathcal{T}\,\varepsilon_{A}\cdot A^{b}\left(x+s\varepsilon_{A}\right)\,\varepsilon_{A}\cdot A^{b'}\left(x+s'\varepsilon_{A}\right)\left|k_{2},\lambda_{2},a_{2}\right\rangle .\end{multline}
Above, $\mathcal{T}$ denotes the time ordering of fields while the
colon the normal ordering of fields.

Let us first calculate $M_{1}$. Using the Wick theorem we get\begin{multline}
M_{1}=-\frac{ig}{2\pi}\, f_{acd}\delta_{c_{A}b}\varepsilon_{A}^{\alpha}\int_{-\infty}^{\infty}ds\int\widehat{dq}\int d^{4}y\, d^{4}w\, q^{\beta}e^{-i\left(y-w\right)\cdot q}\,\\
\Bigg\{ S_{\alpha\nu}^{ab}\left(x-w+s\varepsilon_{A}\right)\left\langle k_{1},\lambda_{1},a_{1}\right|:A_{\beta}^{c}\left(y\right)A^{d\,\nu}\left(y\right):\left|k_{2},\lambda_{2},a_{2}\right\rangle \\
+S_{\alpha\beta}^{bc}\left(x-y+s\varepsilon_{A}\right)\left\langle k_{1},\lambda_{1},a_{1}\right|:A_{\nu}^{a}\left(w\right)A^{d\,\nu}\left(y\right):\left|k_{2},\lambda_{2},a_{2}\right\rangle \\
+S_{\alpha\nu}^{bd}\left(x-y+s\varepsilon_{A}\right)\left\langle k_{1},\lambda_{1},a_{1}\right|:A^{a\,\nu}\left(w\right)A_{\beta}^{c}\left(y\right):\left|k_{2},\lambda_{2},a_{2}\right\rangle \Bigg\},\end{multline}
where \begin{equation}
S_{\mu\nu}^{ab}\left(x\right)=\int\widehat{dp}\, e^{-ip\cdot x}\frac{-i\eta_{\mu\nu}}{p^{2}+i\epsilon}\end{equation}
is a position space Feynman propagator for a gluon. For the subsequent
matrix elements we get explicitly\begin{multline}
\left\langle k_{1},\lambda_{1},c_{1}\right|:A_{\beta}^{c}\left(y\right)A^{d\,\nu}\left(y\right):\left|k_{2},\lambda_{2},c_{2}\right\rangle =\\
=e^{iy\cdot\left(k_{1}-k_{2}\right)}\left[\delta_{cc_{1}}\delta_{dc_{2}}\varepsilon_{\beta}^{\lambda_{1}*}\left(k_{1}\right)\varepsilon^{\lambda_{2}\,\nu}\left(k_{2}\right)\,+\delta_{dc_{1}}\delta_{cc_{2}}\varepsilon_{\beta}^{\lambda_{2}}\left(k_{2}\right)\varepsilon^{\lambda_{1}\,\nu*}\left(k_{1}\right)\right],\end{multline}

\begin{multline}
\left\langle k_{1},\lambda_{1},c_{1}\right|:A_{\nu}^{a}\left(w\right)A^{d\,\nu}\left(y\right):\left|k_{2},\lambda_{2},c_{2}\right\rangle =\delta_{ac_{1}}\delta_{dc_{2}}\varepsilon_{\nu}^{\lambda_{1}*}\left(k_{1}\right)\varepsilon^{\lambda_{2}\,\nu}\left(k_{2}\right)\, e^{iw\cdot k_{1}-iy\cdot k_{2}}\\
+\delta_{dc_{1}}\delta_{ac_{2}}\varepsilon_{\nu}^{\lambda_{2}}\left(k_{2}\right)\varepsilon^{\lambda_{1}\,\nu*}\left(k_{1}\right)\, e^{-iw\cdot k_{2}+iy\cdot k_{1}},\end{multline}
\begin{multline}
\left\langle k_{1},\lambda_{1},c_{1}\right|:A^{a\,\nu}\left(w\right)A_{\beta}^{c}\left(y\right):\left|k_{2},\lambda_{2},c_{2}\right\rangle =\delta_{ac_{1}}\delta_{cc_{2}}\varepsilon^{\lambda_{1}\,\nu*}\left(k_{1}\right)\varepsilon_{\beta}^{\lambda_{2}}\left(k_{2}\right)\, e^{iw\cdot k_{1}-iy\cdot k_{2}}\\
+\delta_{cc_{1}}\delta_{ac_{2}}\varepsilon^{\lambda_{2}\,\nu}\left(k_{2}\right)\varepsilon_{\beta}^{\lambda_{1}*}\left(k_{1}\right)\, e^{-iw\cdot k_{2}+iy\cdot k_{1}}.\end{multline}
Inserting this to $M_{1}$ we get\begin{multline}
M_{1}=-\frac{ig}{2\pi}\, f_{acd}\delta_{c_{A}b}\varepsilon_{A}^{\alpha}\int_{-\infty}^{\infty}ds\int d^{4}qd^{4}p\, q^{\beta}e^{-ip\cdot x}e^{-isp\cdot\varepsilon_{A}}\frac{-i}{p^{2}}\,\\
\Bigg\{\eta_{\alpha\nu}\delta^{4}\left(q+p\right)\delta^{4}\left(q-k_{1}+k_{2}\right)\delta_{ab}\\
\Big[\delta_{cc_{1}}\delta_{dc_{2}}\varepsilon_{\beta}^{\lambda_{1}*}\left(k_{1}\right)\varepsilon^{\lambda_{2}\,\nu}\left(k_{2}\right)+\delta_{dc_{1}}\delta_{cc_{2}}\varepsilon_{\beta}^{\lambda_{2}}\left(k_{2}\right)\varepsilon^{\lambda_{1}\,\nu*}\left(k_{1}\right)\Big]\\
+\eta_{\alpha\beta}\Big[\delta\left(q+k_{1}\right)\delta\left(q-p+k_{2}\right)\delta_{bc}\delta_{ac_{1}}\delta_{dc_{2}}\varepsilon_{\nu}^{\lambda_{1}*}\left(k_{1}\right)\varepsilon^{\lambda_{2}\,\nu}\left(k_{2}\right)\\
+\delta\left(q-k_{2}\right)\delta\left(q-p-k_{1}\right)\delta_{bc}\delta_{dc_{1}}\delta_{ac_{2}}\varepsilon_{\nu}^{\lambda_{2}}\left(k_{2}\right)\varepsilon^{\lambda_{1}\,\nu*}\left(k_{1}\right)\Big]\\
+\eta_{\alpha\nu}\Big[\delta\left(q+k_{1}\right)\delta\left(q-p+k_{2}\right)\delta_{bd}\delta_{ac_{1}}\delta_{cc_{2}}\varepsilon^{\lambda_{1}\,\nu*}\left(k_{1}\right)\varepsilon_{\beta}^{\lambda_{2}}\left(k_{2}\right)\\
+\delta\left(q-k_{2}\right)\delta\left(q-p-k_{1}\right)\delta_{bd}\delta_{cc_{1}}\delta_{ac_{2}}\varepsilon^{\lambda_{2}\,\nu}\left(k_{2}\right)\varepsilon_{\beta}^{\lambda_{1}*}\left(k_{1}\right)\Big]\Bigg\}.\end{multline}
Now, we perform the Fourier transform with respect to $x$ and utilize
the deltas\begin{multline}
\tilde{M}_{1}=-\frac{g}{2\pi}\,\delta^{4}\left(k_{A}+k_{1}-k_{2}\right)\int_{-\infty}^{\infty}ds\, e^{-isk_{A}\cdot\varepsilon_{A}}\frac{\varepsilon_{A}^{\alpha}}{k_{A}^{2}}\, f_{c_{A}c_{1}c_{2}}\\
\Bigg\{-\eta_{\alpha\nu}k_{A}^{\beta}\left[\varepsilon_{\beta}^{\lambda_{1}*}\left(k_{1}\right)\varepsilon^{\lambda_{2}\,\nu}\left(k_{2}\right)-\varepsilon_{\beta}^{\lambda_{2}}\left(k_{2}\right)\varepsilon^{\lambda_{1}\,\nu*}\left(k_{1}\right)\right]\\
+\eta_{\alpha\beta}\left[k_{1}^{\beta}\varepsilon_{\nu}^{\lambda_{1}*}\left(k_{1}\right)\varepsilon^{\lambda_{2}\,\nu}\left(k_{2}\right)+k_{2}^{\beta}\varepsilon_{\nu}^{\lambda_{2}}\left(k_{2}\right)\varepsilon^{\lambda_{1}\,\nu*}\left(k_{1}\right)\right]\\
-\eta_{\alpha\nu}\left[k_{1}^{\beta}\varepsilon^{\lambda_{1}\,\nu*}\left(k_{1}\right)\varepsilon_{\beta}^{\lambda_{2}}\left(k_{2}\right)+k_{2}^{\beta}\varepsilon^{\lambda_{2}\,\nu}\left(k_{2}\right)\varepsilon_{\beta}^{\lambda_{1}*}\left(k_{1}\right)\right]\Bigg\}.\end{multline}
Using\begin{equation}
\int_{-\infty}^{\infty}ds\, e^{-isp\cdot\varepsilon_{A}}=2\pi\,\delta\left(p\cdot\varepsilon_{A}\right)\end{equation}
and rearranging the terms we get\begin{multline}
\tilde{M}_{1}=-g\,\delta^{4}\left(k_{A}+k_{1}-k_{2}\right)\delta\left(k_{A}\cdot\varepsilon_{A}\right)\,\frac{\varepsilon_{A}^{\alpha}\eta_{\alpha\gamma}}{k_{A}^{2}}\, f_{c_{A}c_{1}c_{2}}\\
\left[\eta^{\alpha_{1}\alpha_{2}}\left(k_{1}^{\gamma}+k_{2}^{\gamma}\right)+\eta^{\gamma\alpha_{1}}\left(k_{A}^{\alpha_{2}}-k_{1}^{\alpha_{2}}\right)+\eta^{\alpha_{2}\gamma}\left(-k_{2}^{\alpha_{1}}-k_{A}^{\alpha_{1}}\right)\right]\varepsilon_{\alpha_{1}}^{\lambda_{1}*}\left(k_{1}\right)\varepsilon_{\alpha_{2}}^{\lambda_{2}}\left(k_{2}\right).\end{multline}
Let us note, that in the square bracket we have the standard triple-gluon
vertex.

Next let us calculate $M_{2}$. We have\begin{multline}
M_{2}=-\frac{g}{\pi}\,\mathrm{Tr}\left(t^{c_{A}}t^{b}t^{b'}\right)\varepsilon_{A}^{\alpha}\varepsilon_{A}^{\beta}\int_{-\infty}^{\infty}ds\int_{-\infty}^{s}ds'\\
\left\langle k_{1},\lambda_{1},a_{1}\right|:A_{\alpha}^{b}\left(x+s\varepsilon_{A}\right)\, A_{\beta}^{b'}\left(x+s'\varepsilon_{A}\right):\left|k_{2},\lambda_{2},a_{2}\right\rangle .\end{multline}
Further\begin{multline}
\left\langle k_{1},\lambda_{1},a_{1}\right|:A_{\alpha}^{b}\left(x+s\varepsilon_{A}\right)\, A_{\beta}^{b'}\left(x+s'\varepsilon_{A}\right):\left|k_{2},\lambda_{2},a_{2}\right\rangle =\\
=e^{ix\cdot\left(k_{1}-k_{2}\right)}\Big[e^{is\varepsilon_{A}\cdot k_{1}}e^{-is'\varepsilon_{A}\cdot k_{2}}\delta_{bc_{1}}\delta_{b'c_{2}}\varepsilon_{\alpha}^{\lambda_{1}*}\left(k_{1}\right)\varepsilon_{\beta}^{\lambda_{2}}\left(k_{2}\right)\\
+e^{-is\varepsilon_{A}\cdot k_{2}}e^{is'\varepsilon_{A}\cdot k_{1}}\delta_{bc_{2}}\delta_{b'c_{1}}\varepsilon_{\alpha}^{\lambda_{2}}\left(k_{2}\right)\varepsilon_{\beta}^{\lambda_{1}*}\left(k_{1}\right)\Big].\end{multline}
Thus we have\begin{multline}
M_{2}=-\frac{g}{\pi}\,\mathrm{Tr}\left(t^{c_{A}}t^{b}t^{b'}\right)\varepsilon_{A}^{\alpha}\varepsilon_{A}^{\beta}\int_{-\infty}^{\infty}ds\int_{-\infty}^{s}ds'\, e^{ix\cdot\left(k_{1}-k_{2}\right)}\\
\Big[e^{is\varepsilon_{A}\cdot k_{1}}e^{-is'\varepsilon_{A}\cdot k_{2}}\delta_{bc_{1}}\delta_{b'c_{2}}\varepsilon_{\alpha}^{\lambda_{1}*}\left(k_{1}\right)\varepsilon_{\beta}^{\lambda_{2}}\left(k_{2}\right)\\
+e^{-is\varepsilon_{A}\cdot k_{2}}e^{is'\varepsilon_{A}\cdot k_{1}}\delta_{bc_{2}}\delta_{b'c_{1}}\varepsilon_{\alpha}^{\lambda_{2}}\left(k_{2}\right)\varepsilon_{\beta}^{\lambda_{1}*}\left(k_{1}\right)\Big].\end{multline}
Next, we have to integrate over the path parameters. Using the $i\epsilon$
prescription (\ref{eq:eikonal_prop}) we have\begin{equation}
\int_{-\infty}^{\infty}ds\,\int_{-\infty}^{s}ds'\, e^{is\varepsilon_{A}\cdot k_{1}}e^{-is'\varepsilon_{A}\cdot k_{2}}=\frac{2\pi i}{\varepsilon_{A}\cdot k_{2}+i\epsilon}\delta\left(\varepsilon_{A}\cdot k_{1}-\varepsilon_{A}\cdot k_{2}\right),\end{equation}
\begin{equation}
\int_{-\infty}^{\infty}ds\,\int_{-\infty}^{s}ds'\, e^{-is\varepsilon_{A}\cdot k_{2}}e^{is'\varepsilon_{A}\cdot k_{1}}=\frac{2\pi i}{-\varepsilon_{A}\cdot k_{1}+i\epsilon}\delta\left(\varepsilon_{A}\cdot k_{1}-\varepsilon_{A}\cdot k_{2}\right).\end{equation}
Taking the Fourier transform we get\begin{multline}
\tilde{M}_{2}=-2g\,\delta^{4}\left(k_{A}+k_{1}-k_{2}\right)\varepsilon_{A}\cdot\varepsilon^{\lambda_{1}*}\left(k_{1}\right)\,\varepsilon_{A}\cdot\varepsilon^{\lambda_{2}}\left(k_{2}\right)\\
\delta\left(\varepsilon_{A}\cdot k_{A}\right)\Bigg[\mathrm{Tr}\left(t^{c_{A}}t^{c_{1}}t^{c_{2}}\right)\frac{i}{\varepsilon_{A}\cdot k_{2}}\\
+\mathrm{Tr}\left(t^{c_{A}}t^{c_{2}}t^{c_{1}}\right)\frac{-i}{\varepsilon_{A}\cdot k_{1}}\Bigg].\end{multline}
or\begin{equation}
\tilde{M}_{2}=g\,\delta^{4}\left(k_{A}+k_{1}-k_{2}\right)\varepsilon_{A}\cdot\varepsilon^{\lambda_{1}*}\left(k_{1}\right)\,\varepsilon_{A}\cdot\varepsilon^{\lambda_{2}}\left(k_{2}\right)\frac{1}{\varepsilon_{A}\cdot k_{1}}\, f_{c_{A}c_{1}c_{2}}.\end{equation}
Collecting $\tilde{M}_{1}$ and $\tilde{M}_{2}$ we get\begin{multline}
\mathfrak{M}\left(\varepsilon_{A};\varepsilon_{1},\varepsilon_{2}\right)=-g\,\delta^{4}\left(k_{A}+k_{1}-k_{2}\right)\delta\left(k_{A}\cdot\varepsilon_{A}\right)\, f_{c_{A}c_{1}c_{2}}\varepsilon_{A\gamma}\varepsilon_{\alpha_{1}}^{\lambda_{1}*}\left(k_{1}\right)\varepsilon_{\alpha_{2}}^{\lambda_{2}}\left(k_{2}\right)\\
\Bigg\{\frac{1}{k_{A}^{2}}\left[\eta^{\alpha_{1}\alpha_{2}}\left(k_{1}^{\gamma}+k_{2}^{\gamma}\right)+\eta^{\gamma\alpha_{1}}\left(k_{A}^{\alpha_{2}}-k_{1}^{\alpha_{2}}\right)+\eta^{\alpha_{2}\gamma}\left(-k_{2}^{\alpha_{1}}-k_{A}^{\alpha_{1}}\right)\right]\\
-\frac{\varepsilon_{A}^{\alpha_{1}}\varepsilon_{A}^{\alpha_{2}}}{\varepsilon_{A}\cdot k_{1}}\Bigg\}.\end{multline}
Let us note, that for $\varepsilon_{A}$ taken to be $n_{+}$ or $n_{-}$
and after multiplying by $k_{A}^{2}$ it coincides with the RPP vertex
of \cite{Antonov:2004hh}.

\section{The BRST invariance of the $\mathcal{R}$ operator}

\label{sec:BRST_R}

In this appendix we argue that the state (\ref{eq:Rstate}) defined
by the action of the operator (\ref{eq:Rphys_def}) on the vacuum
state belongs to the cohomology of the BRST transformation. For a
pedagogical review of the subject we refer to \cite{Weinberg:1996kr}.
Here, let us just recall the necessary basic facts. Let $Q_{\mathrm{BRST}}$
be a charge generating the BRST transformation. It is nilpotent, i.e.
$Q_{\mathrm{BRST}}^{2}=0$. Below, the following commutation relation
will be useful: $i\left[Q_{\mathrm{BRST}},A_{\mu}^{a}\right]=D_{\mu}^{ab}c^{b}$,
where $D_{\mu}^{ab}=\delta^{ab}\partial_{\mu}-gf^{abc}A_{\mu}^{c}$
and $c^{b}$ is the ghost field. A state $\left|\phi\right\rangle $
belongs to the cohomology of $Q_{\mathrm{BRST}}$ if it belongs to
the kernel of $Q_{\mathrm{BRST}}$ (i.e. $Q_{\mathrm{BRST}}\left|\phi\right\rangle =0$)
but not to the image. From nilpotency it directly follows that any
state in the image has zero norm. It is the basic requirement for
physical states that they belong to the cohomology of the BRST transformation.

Let us now turn to the $\mathcal{R}_{n}^{c}$ operator. Note that
since the vacuum state belongs to the cohomology and since the state
$\mathcal{R}_{n}^{c}\left|0\right\rangle $ has a nonzero norm, it
is enough to show that \begin{equation}
\left[\mathcal{R}_{n}^{d}\left(k\right),Q_{\mathrm{BRST}}\right]=0.\label{eq:BRST_R_commutator}\end{equation}
To this end we expand the definition (\ref{eq:Rphys_def}) of $\mathcal{R}_{n}^{c}$.
The first nontrivial term of the expansion of (\ref{eq:BRST_R_commutator})
is\begin{multline}
i\int\frac{dk^{\left(n\right)}}{2\pi}\int d^{4}x\, e^{ix\cdot k}\int_{-\infty}^{\infty}ds\, n^{\mu}t^{a}\left[Q_{\mathrm{BRST}},A_{\mu}^{a}\left(x+sn\right)\right]\\
=-i\int dk^{\left(n\right)}\delta\left(k\cdot n\right)k\cdot n\,\, t^{a}\tilde{c}^{a}\left(k\right)\\
-g\int dk^{\left(n\right)}\int\frac{d^{4}p}{\left(2\pi\right)^{4}}\,\delta\left(k\cdot n\right)t^{a}f^{abc}n^{\mu}\tilde{A}_{\mu}^{c}\left(p\right)\tilde{c}^{b}\left(k-p\right),\label{eq:BRST_R_exp1}\end{multline}
where we have switched to the Fourier space (the tildes denote Fourier-transformed
fields). Note that the first term on the r.h.s vanishes after integrating
the delta function. The second term of the expansion of the commutator
(\ref{eq:BRST_R_commutator}) is\begin{multline}
i^{2}g\int\frac{dk^{\left(n\right)}}{2\pi}\int d^{4}x\, e^{ix\cdot k}\int_{-\infty}^{\infty}ds\int_{-\infty}^{s}ds'\, n^{\mu}n^{\nu}t^{a}t^{a'}\left[Q_{\mathrm{BRST}},A_{\mu}^{a}\left(x+sn\right)A_{\nu}^{a'}\left(x+s'n\right)\right]\\
=g\int dk^{\left(n\right)}\int\frac{d^{4}p}{\left(2\pi\right)^{4}}\frac{d^{4}p'}{\left(2\pi\right)^{4}}\delta^{4}\left(k-p-p'\right)\delta\left(p\cdot n+p'\cdot n\right)p'\cdot n\\
\left\{ \frac{i}{p'\cdot n+i\epsilon}t^{a}t^{a'}+\frac{i}{p\cdot n+i\epsilon}t^{a'}t^{a}\right\} n^{\mu}\tilde{A}_{\mu}^{a}\left(p\right)c^{a'}\left(p'\right)+\mathcal{O}\left(g^{2}\right)\\
=ig\int dk^{\left(n\right)}\int\frac{d^{4}p}{\left(2\pi\right)^{4}}\delta\left(k\cdot n\right)\left[t^{a},t^{a'}\right]n^{\mu}\tilde{A}_{\mu}^{a}\left(p\right)\tilde{c}^{a'}\left(k-p\right)+\mathcal{O}\left(g^{2}\right),\label{eq:BRST_R_exp2}\end{multline}
where we have suppressed the term of the order of $g^{2}$. Using
$\left[t^{a},t^{a'}\right]=if^{aa'b}t^{b}$ and reshuffling the indices
we see that the first term on the r.h.s cancels against (\ref{eq:BRST_R_exp1}).
Similar canceling will occur order by order in $g$.

\bibliographystyle{JHEP}
\bibliography{small_x}

\end{document}